\documentclass[twocolumn,showpacs,preprintnumbers,amsmath,amssymb]{revtex4}

\usepackage{subfigure,dcolumn,jabbrv}
\usepackage{graphicx}
\usepackage{dcolumn}
\usepackage{bm}
\usepackage{CJK}
\usepackage{url}
\usepackage{color}
\usepackage{booktabs}
\usepackage{natbib}
\usepackage{hyperref}
\usepackage{cleveref}

\begin{document}

\title{Evolutions of in-medium baryon-baryon scattering cross sections and stiffness of dense nuclear matter from Bayesian analyses of FOPI proton flow excitation functions}



\author{Bao-An Li$^{1}$$\footnote{Corresponding Author: Bao-An.Li@Tamuc.edu}$,  Wen-Jie Xie$^{1,2,3,4}$$\footnote{wenjiexie@yeah.net}$}
\affiliation{$^1$Department of Physics and Astronomy, East Texas A$\&$M University, Commerce, TX 75429-3011, USA}
\affiliation{$^{2}$Shanxi Province Intelligent Optoelectronic Sensing Application Technology Innovation Center, Yuncheng University, Yuncheng 044000, China}
\affiliation{$^3$Guangxi Key Laboratory of Nuclear Physics and Nuclear Technology, Guangxi Normal University, Guilin 541004, China}
\affiliation{$^4$Shanxi Province Optoelectronic Information Science and Technology Laboratory, Yuncheng University, Yuncheng 044000, China}
\date{\today}

\begin{abstract}
Within a Bayesian statistical framework using a Gaussian Process (GP) emulator for an isospin-dependent Boltzmann-Uehling-Uhlenbeck (IBUU) transport model simulator of heavy-ion reactions with momentum-independent Skyrme interactions, we infer from the proton directed and elliptical flow in mid-central Au+Au reactions at beam energies from 150 to 1200 MeV/nucleon taken by the FOPI Collaboration the posterior Probability Distribution Functions (PDFs) of the in-medium baryon-baryon scattering cross section (BBSCS) modification factor $X$ (with respect to their free-space values) and the stiffness parameter $K$ of dense nuclear matter. We find that the most probable value of $X$ evolves from around 0.7 to 1.0 as the beam energy $E_{beam}/A$ increases. On the other hand, the posterior PDF($K$) may have dual peaks having roughly the same height or extended shoulders at high $K$ values. More quantitatively, the posterior PDF($K$) changes from having a major peak around 220 MeV characterizing a soft EOS in the reaction at $E_{beam}/A$=150 MeV to one that peaks around 320 MeV indicating a stiff EOS in the reactions at $E_{beam}/A$ higher than about 600 MeV. The transition from soft to stiff happens in mid-central Au+Au reactions at beam energies around 250 MeV/nucleon in which $K=220$ MeV and $K=320$ MeV are approximately equally probable. Altogether, the FOPI proton flow excitation function data indicate a gradual hardening of hot and dense nuclear matter as its density and temperature increase in reactions with higher beam energies.

\end{abstract}



\maketitle
\section{Introduction}\label{S1}
Various components of nuclear collective flow in heavy-ion collisions \cite{pawel85,oll,art}
have long been well known for their abilities to reveal the pressure gradient generated, thus providing useful information about both the Equation of State (EOS) and transport properties (e.g, viscosity or in-medium particle-particle scattering cross sections) of hot and dense nuclear matter created transiently during the reactions \cite{Sto86,Bertsch,Cas90,res97,Bass,Pawel02,Buss,Heinz13,QCD,Trautmann,Sor24,Rus23,Wang20}. More specifically, in the local rest 
frame of matter the collective flow velocity $\vec v$, energy density $e$ 
and pressure $P$ are related by the relativistic Euler equation
\begin{equation}
(e+P)\frac{\partial}{\partial t}\vec v=-\vec{\nabla}P.
\end{equation}
In essence, the pressure gradient $-\vec{\nabla}P$ plays the role of a driving 
force for the acceleration of a quasi-particle of effective mass $(e+P)$ (enthalpy) flowing with a velocity $\vec v$ \cite{pawel}. 
The driving force is determined by the product of the density derivative of pressure $P$ (i.e. the speed of sound squared) and the spacial gradient of density depending on the impact parameter and beam energy of the reaction, while the effective mass of the quasi-particle is determined by the EOS at finite temperature depending on the in-medium particle-particle scattering cross sections. Interestingly, all of them are strongly affected by a complicated interplay of nuclear EOS and transport properties of hot and dense matter formed during heavy-ion reactions. 
\begin{figure}[thb]
\centering
 \resizebox{0.5\textwidth}{!}{
\includegraphics[width=1.\textwidth]{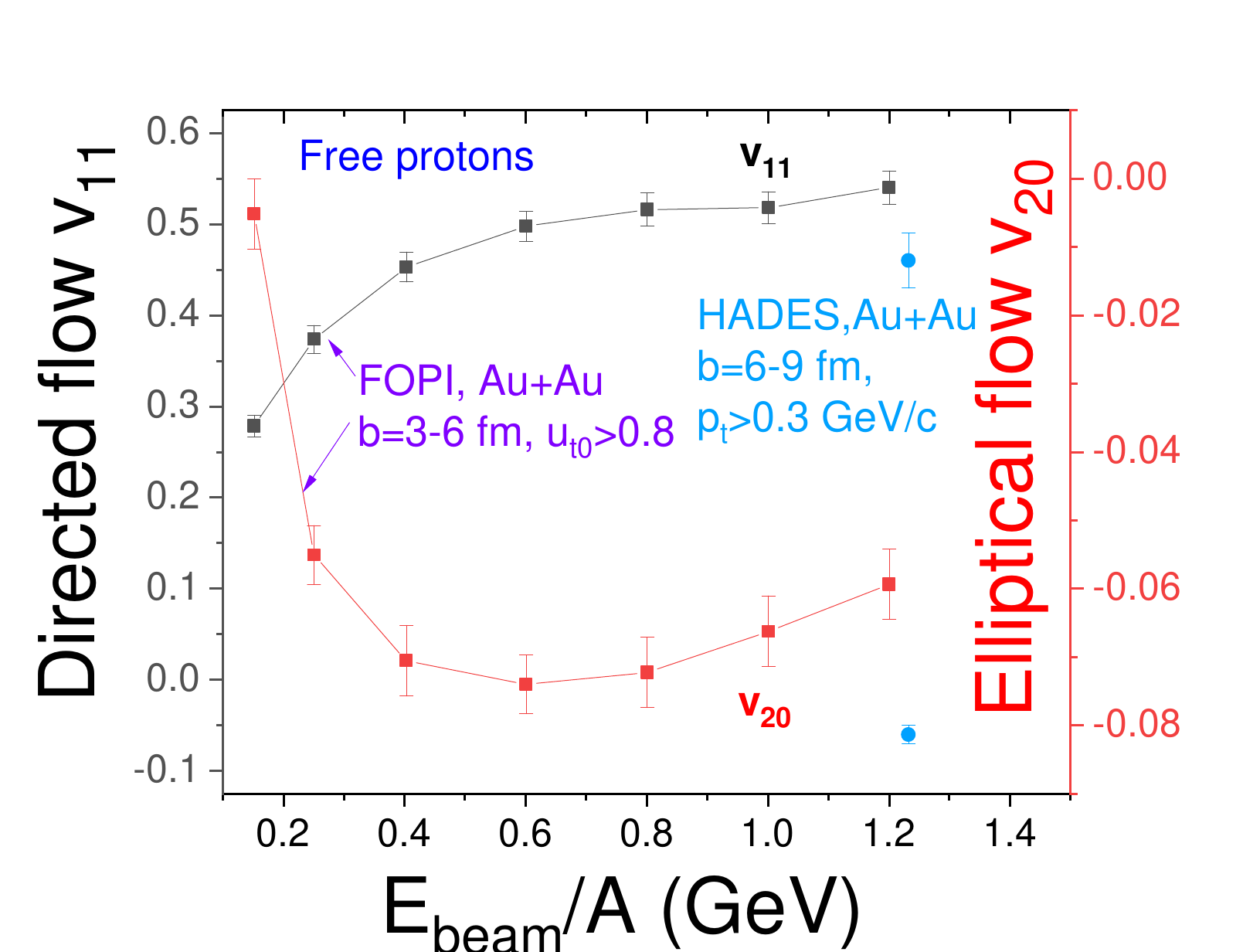}
}
\caption{Excitation functions of proton mid-rapidity slope of directed flow (left axis) and strength of elliptic flow (right axis) in mid-central Au+Au reactions from the FOPI Collaboration \cite{FOPI}. The results at a beam energy of 1.23 GeV/nucleon from the HADES Collaboration \cite{Ha2} with somewhat different centrality and transverse momentum cuts are also shown as a reference.
} \label{v12}
\end{figure}

Indeed, many analyses of experimental data of different flow components carried by various particles in nuclear reactions induced by nucleons, light and heavy ions in a broad energy range from a few tens of MeV/nucleon to LHC energies using hydrodynamics and/or transport models have been very fruitful 
\cite{Sto86,Bertsch,Cas90,res97,Bass,Pawel02,Buss,Heinz13,QCD,Trautmann,Sor24,Rus23,Wang20,pawel}. Nevertheless, many interesting issues remain to be investigated more thoroughly and systematically \cite{EOSBook,LRP1,LRP2,LRP3}. For example, in heavy-ion reactions at intermediate beam energies before QGP is expected to be formed, a longstanding issue is the interplay of stiffness of nuclear EOS (normally characterized by the incompressibility $K$ of symmetric nuclear matter at saturation density $\rho_0$) and the in-medium baryon-baryon scattering cross sections (BBSCSs). The latter is often measured by using the in-medium to free-space cross section ratio  $X\equiv\sigma^{med}_{NN}/\sigma^{free}_{NN}$. There is a well-known $X-K$ degeneracy in describing flow observables, namely the same data can be equally well explained by varying either $K$ or $X$ within their currently known uncertain ranges, 
see, e.g., Refs.
\cite{Bertsch:1988xu,Xu:1991zz,Westfall:1993zz,Klakow:1993dj,TLi93,Bonasera,Alm:1995chb,BALI1,Zheng99,LiSustich,Danielewicz:2002he,Lopez14,BALI2,Tsang24}. 

Presently, it is still not clear how this $X-K$ degeneracy (or relative importance of $K$ and $X$) in describing nuclear collective flow evolves with the beam energy. Uncovering this evolution (checking if the $X-K$ degeneracy is broken) from Bayesian analyses of the proton flow excitation functions measured by the FOPI Collaboration at GSI is the main purpose of this work.  Shown in Fig. \ref{v12}
are the FOPI proton directed flow $v_{11}=F_1\equiv dv_1/dy'|_{y'=0}$ (left axis) and elliptical flow $v_{20}$ at mid-rapidity (right axis) as functions of beam energy in mid-central Au+Au reactions, respectively. More detailed definitions of the quantities involved are given in Section \ref{IBUU-P}. Among all flow data available in reactions with beam energies ranging from tens of MeV to several TeV per nucleon \cite{Ha2,EOSBook,LRP1}, the FOPI experiments measured proton directed and elliptic flows systematically in a broad beam energy range where the flow data show some very interesting features. In particular, the proton directed and elliptical flows are anti-correlated and have much larger magnitudes than reactions at higher beam energies available at RHIC and LHC. Digging out novel information about the EOS and transport properties of hot and dense hadronic matter from the FOPI data has been very appealing. While many analyses of the complete or partial FOPI data within various transport models have been done already, see, e.g., Refs. \cite{FOPI,f1,f2,f3,f4,f5,f6,f7,f8,f9,f10,PLi18,Li:2022wvu,f11,f12,f13}, they are mostly forward-modelings in the traditional approaches. We report here results of our Bayesian analyses of the FOPI proton flow excitation function data. We found strong indications that the EOS of hot and dense matter created in mid-central Au+Au reactions becomes harder as its density and temperature increase with beam energy from 150 to 1200 MeV/nucleon.

The rest of the paper is organized as follows: in the next section we briefly summarize the IBUU modeling and Bayesian analysis approach. In particular, we discuss how the single parameter $K$ uniquely characterizes the stiffness (measured by the speed of sound) of dense nuclear matter with the momentum-independent Skyrme interaction used in the present work. In Section \ref{Results}, we present and discuss our results. In particular, in subsection \ref{IBUU-P}, using the IBUU in the traditional approach we investigate the individual roles of the in-medium BBSCS and the stiffness of nuclear EOS on the evolution of proton directed and elliptical flow from mid-central Au+Au reactions with a beam energy from 150 MeV to 1200 MeV per nucleon. These data are then used as training and validating data for the GP emulator of the IBUU simulator. In subsection \ref{B-results}, we discuss results of our Bayesian inference of the 
evolution of the in-medium BBSCS and the stiffness of nuclear EOS. Finally, we summarize our work in Section \ref{sum}.
Though out the paper, we use the abbreviation PDF for Probability Distribution Functions.

\section{Main ingredients of IBUU modeling and Bayesian inference}\label{approaches}
In our recent work in Ref. \cite{LX-HA} within a comprehensive IBUU+GP+Bayesian framework, the posterior PDFs of in-medium BBSCS modification factor $X$ and incompressibility $K$ were inferred from HADES proton flow data  \cite{Ha2} shown in Fig. \ref{v12} using a GP emulator for the IBUU transport model simulator \cite{BALI1,BALI,libauer2,LiBA04} employing a momentum-independent Skyrme interaction. A mean value of $X=1.32^{+0.28}_{-0.40}$ and $K=346^{+29}_{-31}$ MeV were found, indicating an enhanced in-medium BBSCS with respect to their free-space values and a rather stiff nuclear EOS. In this work, using the same IBUU+GP+Bayesian approach we infer the posterior PDFs of $X$ and $K$ as functions of beam energy from the FOPI excitation functions of proton directed and elliptical flow. For completeness and ease of our discussions below, we briefly recall here a few key elements of our approach. More details can be found in our previous publication \cite{LX-HA}.
To illustrate how the single parameter $K$ in Skyrme energy density functionals can describe completely the stiffness of cold nuclear matter in a broad density range far above the saturation density $\rho_0$ without violating causality, we add new discussions here on the correlations among the curvature, skewness and kurtosis as well as the density dependence of speed of sound $C^2_S(\rho)$. The latter can be used to measure the overall stiffness of cold nuclear matter and it is uniquely characterized by $K$ within the Skyrme energy density functionals. 

\begin{figure*}[thb]
\centering
 \resizebox{1.\textwidth}{!}{
\includegraphics[width=1.\textwidth]{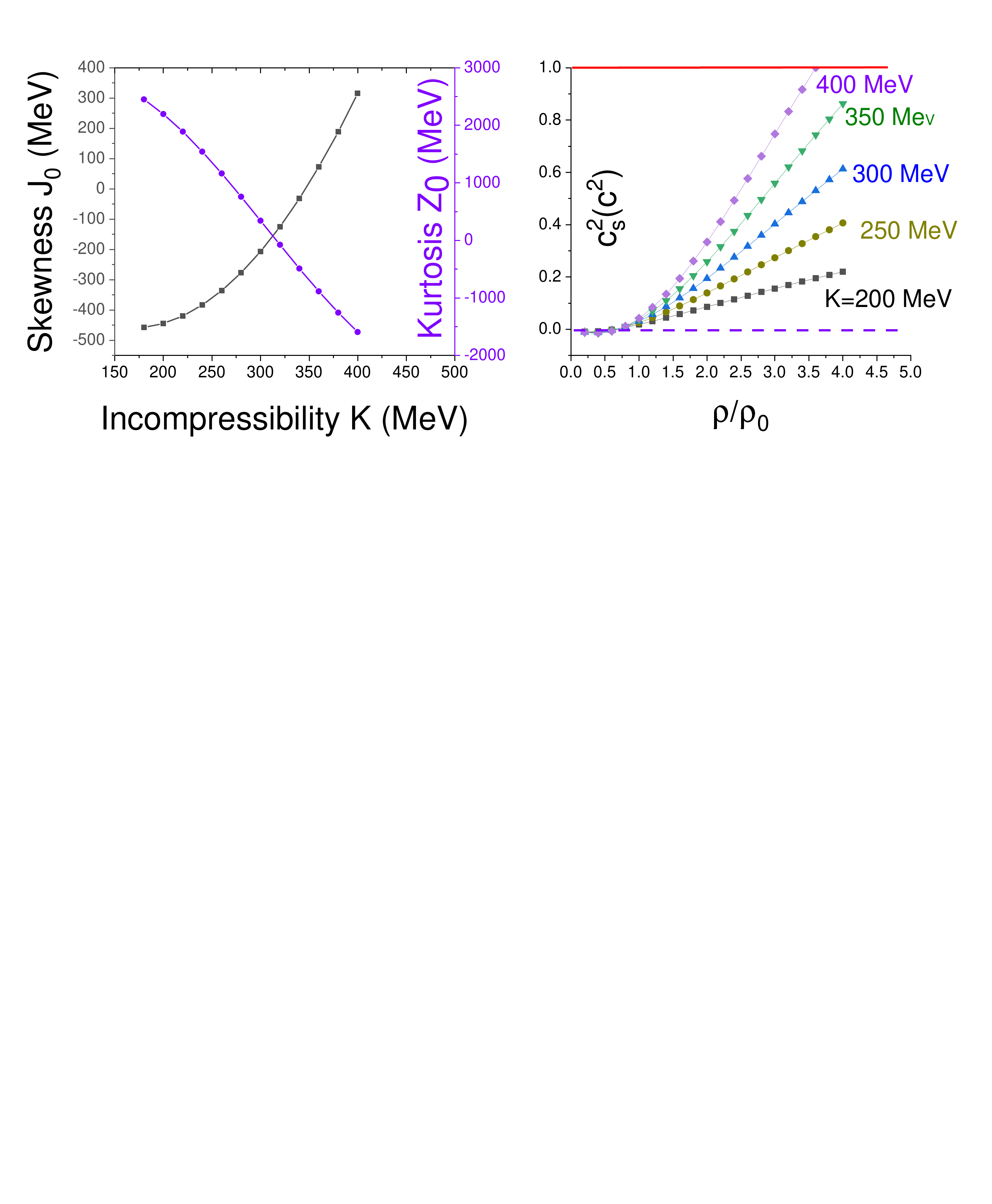}
}
\vspace{-14cm}
\caption{Left: The skewness $J_0$ and kurtosis $Z_0$ as functions of incompressibility $K$. Right: the adiabatic speed of sound squared as a function of density with $K$ between 200 and 400 MeV using the Skyrme energy density functions given in the text.
} \label{J0Z0K}
\end{figure*}
\subsection{EOS used in the IBUU simulator}
The IBUU simulator has multiple choices for the baryon mean-field potential and in-medium BBSCS \cite{BALI}. Because of the limited computing power we have and the task to generate the necessarily large data sets to train and test the GP emulator of the IBUU simulator, we use the following Skyrme-type momentum-independent potential for baryon $q$
\begin{equation}\label{vq}
      V_{q}(\rho,\delta) = a (\rho/\rho_0) + b (\rho/\rho_0)^{\sigma}\
	+V_{\rm asy}^{q}(\rho,\delta) +V^{q}_{\rm Coulomb}.
\end{equation}
The $V_{\rm asy}^{q}(\rho,\delta)$ is the baryon symmetry potential in asymmetric nuclear matter with isospin asymmetry $\delta$. Here we adopt (corresponding to the $F_3$ in Eq. (3) of Ref. \cite{Li97}) the following momentum-independent symmetry potential
\begin{equation}
V_{\rm asy}^{n(p)}=\pm 2e_a (\rho/\rho_0)^{1/2}\delta-\frac{1}{2}e_a(\rho/\rho_0)^{1/2}\delta^2
\end{equation}
where $e_a\equiv E_{\rm sym}(\rho_0)-(2^{2/3}-1)\,{\textstyle\frac{3}{5}}E_F^0$ with the symmetry energy $E_{\rm sym}(\rho_0)=32$ MeV and the Fermi energy $E_F^0=36$ MeV at $\rho_0$. The $\pm$ sign is for neutrons/protons, respectively, and $V^{q}_{\rm Coulomb}$ is the Coulomb potential for charged particles. The potentials of baryon resonances ($\Delta$ and $N^*$) are related to those of nucleons through the square of the Clebsch-Gordon coefficients in their decays to pion+nucleon processes \cite{Linpa}.

The three parameters $a,~b$ and $\sigma$ in Eq. (\ref{vq}) are determined by the
saturation properties (binding energy of -16 MeV and vanishing pressure at $\rho_0=0.16$ fm$^{-3}$) as well as a specific incompressibility $K$ at $\rho_0$ 
\begin{equation}
K=9\rho_0^2[d^2E_0(\rho)/d \rho^2]_{\rho_0}
\end{equation}
where $E_0(\rho)$ is the energy per nucleon in cold symmetric nuclear matter at density $\rho$. More specifically, we have \cite{Bertsch,BALI1}
\begin{eqnarray}
a&=&-29.81-46.90\frac{K+44.73}{K-166.32}~({\rm MeV}),\\
b&=&23.45\frac{K+255.78}{K-166.32}~({\rm MeV}),\\
\sigma&=&\frac{K+44.73}{211.05}.
\end{eqnarray} 
Thus, once the single free parameter $K$ characterizing the EOS is specified, the entire density dependence of the EOS is determined. For example, the skewness and kurtosis at $\rho_0$ 
\begin{equation}
J_0=27\rho_0^3[d^3E_0(\rho)/d \rho^3]_{\rho_0}~{\rm and}~Z_0=81\rho_0^4[d^4E_0(\rho)/d \rho^4]_{\rho_0}
\end{equation}
are then also determined completely. For an illustration, they are shown in the left panel of Fig. \ref{J0Z0K} as functions of $K$. It is seen that while the $K$ and $J_0$ are positively correlated, the $K$ and $Z_0$ are anti-correlated. While in calculating $J_0$ and $Z_0$ the high-order derivatives of $E_0(\rho)$ are also taken at $\rho_0$, they actually characterize the density dependence of $E_0(\rho)$ at both sub-saturation and supra-saturation densities significantly away from $\rho_0$. At an arbitrary density $\rho$, the skewness and kurtosis can be written respectively as 
\begin{equation}
      J_{\rho}=27\left[\frac{8}{45}E_F^0(\frac{\rho}{\rho_0})^{2/3}
      +\frac{b\sigma(\sigma-1)(\sigma-2)}{1+\sigma}(\frac{\rho}{\rho_0})^{\sigma}
	\right],
\end{equation}
\begin{equation}
Z_{\rho}=81\left[-\frac{56}{135}E_F^0(\frac{\rho}{\rho_0})^{2/3}
      +\frac{b\sigma(\sigma-1)(\sigma-2)(\sigma-3)}{1+\sigma}(\frac{\rho}{\rho_0})^{\sigma}
	\right].
\end{equation}
They characterize the strength (via the parameter b) and density dependence (via the parameter $\sigma$) of the effective three-body force independently of the two-body force (via the parameter a). Analytically, since all three parameters (a, b and $\sigma$) are completely determined once the K parameter is specified at $\rho_0$, the high-density stiffness of nuclear matter affected by both $J_{\rho}$ and $Z_{\rho}$ is then also determined. 

The density dependence of the overall stiffness of symmetric nuclear matter at zero temperature can be measured by its adiabatic speed of sound squared (SSS) in unit of $c^2$ \cite{Blaizot,DongLai,Jerome22}
\begin{equation}\label{cnm}
C^2_S=\frac{dP/d\rho}{M_N+E(\rho)+P/\rho}.
\end{equation}
For the Skyrme interactions considered, we have the energy per nucleon 
\begin{equation}
      E(\rho) =\frac{3}{5}E_F^0(\frac{\rho}{\rho_0})^{2/3}+\frac{a}{2}\frac{\rho}{\rho_0}+\frac{b}{1+\sigma}(\frac{\rho}{\rho_0})^{\sigma},
\end{equation}
and the ratio of pressure over density
\begin{equation}
    \frac{P}{\rho} =\frac{2}{5}E_F^0(\frac{\rho}{\rho_0})^{2/3}+\frac{a}{2}\frac{\rho}{\rho_0}+\frac{b\sigma}{1+\sigma}(\frac{\rho}{\rho_0})^{\sigma}.
\end{equation}
Shown in the right panel of Fig. \ref{J0Z0K} is the density dependence of $C^2_s(\rho)$ for the incompressibility $K$ between 200 and 400 MeV. For a given $K$ value, the $C^2_S$ continuously increase in the whole density range considered. Even with an extremely large $K$ value of 400 MeV, the $C^2_S$ remains less than 1 (indicating that the dense matter remains causal) up to the maximum density of about $3.5\rho_0$ reached in central Au+Au reactions at a beam energy of about 1200 MeV/nucleon. In short, the evolution of the stiffness of nuclear matter with increasing density is completely determined by $K$ although the latter is mathematically defined as the energy curvature at $\rho_0$. Namely, using the single parameter K is enough to characterize accurately the stiffness of entire Skyrme EOS in a broad density range. We also notice by passing that the $C^2_s(\rho)$ becomes slightly negative at sub-saturation densities below about $0.5\rho_0$ with some of the $K$ values as indicated in Fig. \ref{J0Z0K}. It indicates the onset of spinodal decomposition due to the mechanical instability of dilute nuclear matter \cite{B-Siemens,Chomaz}, see the detailed discussions in Ref. \cite{mec} using the same Skyrme interactions given above.

It is worth noting that in an earlier study involving one of us in Ref. \cite{Li01} on the speed of sound in hot and dense matter formed in heavy-ion reactions, it was found that the SSS is dominated by contributions of both symmetric nuclear matter EOS discussed above and the kinetic pressure. The latter depends on the temperature that is strongly affected by the in-medium BBSCS during heavy-ion reactions. On the other hand, the symmetry energy contribution is negligible at densities less than about $3\rho_0$ mostly because the isospin asymmetry reached is very small. In simulating intermediate energy heavy-ion reactions, the stiffness of hot and dense matter formed during the reactions is thus expected to be mostly determined by the parameters $K$ and $X$.

\subsection{The prior ranges of model parameters $X$ and $K$}\label{Pri}
We use broad prior ranges of $X = 0.3\sim 2$ and $K = 180\sim 400$ MeV based on the following consideration:
\begin{enumerate}
    \item 
It is well known that the momentum-dependence of nuclear potential affects nuclear collective flow in heavy-ion collisions, see, e.g., Refs.  \cite{Aich87,Gale87,Prakash,Gerd,Pan-Pawel,Cozma}. Nevertheless, the same flow data and generally all observables can be equally well described by using a smaller $K$ with momentum dependence or a larger $K$ without it, see, e.g., Ref. \cite{Fuchs06} for a review. Namely, there is a degeneracy between the repulsive density-dependent force (e.g., three-body force) and the repulsive momentum-dependent nucleon optical potential. Since we do not use the momentum-dependent potential in the present study, it is necessary to generate randomly with an equal probability $K$ values within a large range. The prior range of $180-400$ MeV for $K$ is consistent with available information in the literature. 
Nevertheless, we caution that part of the stiffness of dense matter extracted in our analyses here may come from the increasingly more repulsive single-particle potential due to its momentum dependence. 
\item 
The parameter $X$ measures medium effects on the BBSCS. Instead of introducing more parameters to distinguish possibly different medium effects for elastic and inelastic scatterings and their energy dependence, we use the parameter $X$ to modify all free-space experimental nucleon-nucleon scattering cross sections used as default in the original IBUU code. As discussed in more detail in Ref. \cite{LX-HA}, nuclear medium can affect the $X$ value through several factors \cite{Haar,Herman1}: e.g.,  (1) effective masses of baryons due to the momentum dependence of their mean-field potentials \cite{Gale,LiChen05,Li:2005iba}, (2) the interaction matrix element itself due to, e.g., off-shell or collectivity of exchanged mesons \cite{Bertsch:1988xu,Li:1993rwa,Lom96,Fuchs01}, (3) the Pauli blocking of both intermediate and final states \cite{Carlos}, (4) technical issues in modeling baryon-baryon scatterings in transport models \cite{Herman}. Presently, there is no community consensus on the in-medium cross sections. 
While some calculations predicted $X\leq 1$ with some density dependence, others have shown that $X$ can be much larger than 1. 
For example, considering the pion collectivity in dense matter in evaluating the in-medium interaction matrix element, inelastic nucleon-nucleon cross sections in dense medium were found to increase significantly with respect to their free-space values \cite{Bertsch:1988xu}. Practically, with few exceptions (e.g., Ref.  \cite{Zhang:2007gd}), many previous studies have found strong indications for $X\leq 1$ from analyzing the collective flow and nuclear stopping power of heavy-ion collisions at beam energies below about 800 MeV/nucleon, see, e.g., Refs. \cite{PLi18,Li:2022wvu} and references therein. Here, without any presumption about any density dependence of $X$ by generating randomly $X$ between 0.3 and 2.0 uniformly, we investigate how the poster PDF of $X$ evolves from Bayesian analyses of the FOPI excitation functions of proton directed and elliptical flow. 
\end{enumerate}
We notice that to be computationally efficient, it is not necessary to use exactly the same prior ranges at all beam energies. However, this has to be found out first by trials. For example, at low beam energies, the required $X$ and $K$ values to reproduce the flow data within, e.g., $1\sigma$ errors, are not as higher as those necessary at high-energies. As we shall show and discuss in more detail, to minimize the biases we started from using the same prior ranges at all beam energies. 
At the end, it was sufficient to use $X=0.3\sim 1.5$ and $K=180-380$ MeV at $E_{beam}/A=150$ MeV
and all other reactions with $X=0.5\sim 2.0$ and $K=180-400$ MeV. Of course, knowing the IBUU predictions relative to the data after the present calculations, future studies may use reduced prior ranges for both $X$ and $K$. We also emphasize that interpretations of posterior PDFs from our Bayesian analyses have to be done with respect to the uniform prior PDFs used in the ranges given above.

\subsection{Gaussian Process emulation of proton directed and elliptical flow predicted by the IBUU simulator}
Since multi-million Markov chain Monte Carlo (MCMC) steps are needed in the Bayesian analyses of the flow data at each beam energy, we use here the widely used GP emulator with the Squared Exponential (SE) covariance function \cite{GP}. To train and validate it, large sets of IBUU simulations have to be carried out first. Specifically, we generated 120 sets of $X$ and $K$ values on the Latin Hyperlattice \cite{LH} at each beam energy. The standard validation methods \cite{Pratt,Scott,Bass2,Bass3,Ohio} as done in our previous work \cite{LX-HA} are used in testing the GP accuracy.

In generating the IBUU training, validating and testing data, for each set of the X and K parameters we use 200 testparticles/nucleon and generate randomly 100 impact parameters $b$ with the probability density $P(b)\propto b$ between b=3 fm and 6 fm. This range corresponds to approximately the FOPI centrality range for the reduced impact parameter $b_0\equiv b/b_{\rm{max}}$ between 0.25 and 0.45 where $b_{\rm{max}}$ is the maximum value of $b$ for Au+Au reactions \cite{FOPI}. Thus, for each X and K parameter set, 20,000 quasi-independent IBUU events of Au+Au collisions are used in calculating the proton directed and elliptical flow in each step of MCMC process in our Bayesian analyses.
We notice that the FOPI Collaboration has published their results for different centrality bins with various cuts on the transverse momentum for protons and several light ions. In this work, we focus on analyzing their data for free protons shown in Fig.\ref{v12} (taken from Fig. 14 for directed flow and Fig. 29 for elliptical flow in Ref. \cite{FOPI}) in the centrality and transverse momentum bins where the proton flows are the strongest. 

\subsection{IBUU+GP+Bayesian statistical framework}
For completeness and introducing the terminologies, we recall here the Bayesian theorem
\begin{equation}\label{Bay1}
P({\cal M}|D) = \frac{P(D|{\cal M}) P({\cal M})}{\int P(D|{\cal M}) P({\cal M})d\cal M},
\end{equation}
where the denominator is a normalization constant. The $P({\cal M}|D)$ represents the posterior PDF of the model $\cal M$ given the data set $D$. The $P(D|{\cal M})$ is the likelihood function obtained by comparing predictions of the model $\cal M$ with the data $D$, while the $P({\cal M})$ is the prior PDF of the model $\cal M$. In the present study,
M represents the model parameters $X$ and $K$, and $D$ represents the proton directed flow ($v_{11}$) and elliptical flow ($v_{20}$) data from the FOPI Collaboration. We use the standard Gaussian likelihood function and uniform prior PDFs for $X$ and $K$, see Ref. \cite{LX-HA} for more details.

The likelihood function $P(D|{\cal M})$ is calculated from the product of two Gaussian functions comparing the experimentally observed (obs) and theoretically calculated (th) observables ($R_1=v_{11}$ and $R_2=v_{20}$)
\begin{equation}\label{Likelihood-R}
 P(D|{\cal M})=\prod_{j=1}^{2}\frac{1}{\sqrt{2\pi}\sigma_{\mathrm{obs},j}}\exp[-\frac{(R_{\mathrm{th},j}-R_{\mathrm{obs},j})^{2}}{2\sigma_{\mathrm{obs},j}^{2}}],
\end{equation}
where $\sigma_{\mathrm{obs},j}$ represents the uncertainty associated with the observable $R_j$. In principle, the latter should include both statistical and systematical errors of the experimental data and theoretical calculations as well as emulation uncertainties. Presently, we have little knowledge about the systematic errors of the experimental data, have practically no reliable way to estimate systematical errors of the model calculations without introducing dubious assumptions about correlations of theoretical systematic errors and issues related to model dependence of calculations using different transport models. Practically, in similar Bayesian analyses of flow data at RHIC and/or LHC energies, one 
often uses a constant $\sigma_{\mathrm{obs},j}$ ranging from a pessimistic choice ((6-12)\% of the experimental mean value) to a more optimistic one ((3-6)\%), see, e.g., refs. \cite{Scott,Bass2}. These choices account for the typical experimental error of (3–5)\% in flow measurement plus some additional uncertainties. To use the maximum constraining power of the FOPI data itself and see the most optimal posterior PDFs of X and K that one can possibly infer from the data, we did not use any theoretical error (statistical or systematic) but the minimum 
value of $\sigma_{\mathrm{obs},j}$ given by the experimental statistical error. The results presented here provide the best-case scenario, albeit at the risk of being overly confident but not ignorant. Thus, readers should understand our results within the context given above.  

In our MCMC sampling of posterior PDFs we use the Metropolis-Hastings algorithm. The running averages of model parameters 
and the $-\rm{Log}(\rm{Likelihood~ Function})$ as functions of MCMC steps are used to check if and when the MCMC has reached the equilibrium stage before we start accumulating posterior data. We shall present results from analyzing combined 10 independent MCMC walkers after each has gone through 1 million post burn-in steps. As normally done in the literature, besides analyzing the two-dimensional (2D) poster PDFs (also referred as correlations), we study the so-called marginalized one-dimensional poster PDFs by integrating the 2D PDFs over one of the two model parameters $X$ and $K$. 
\begin{figure}[thb]
\vspace{-1.2cm}
\hspace{-1.2cm}
\centering
 \resizebox{0.55\textwidth}{!}{
\includegraphics[width=1.\textwidth]{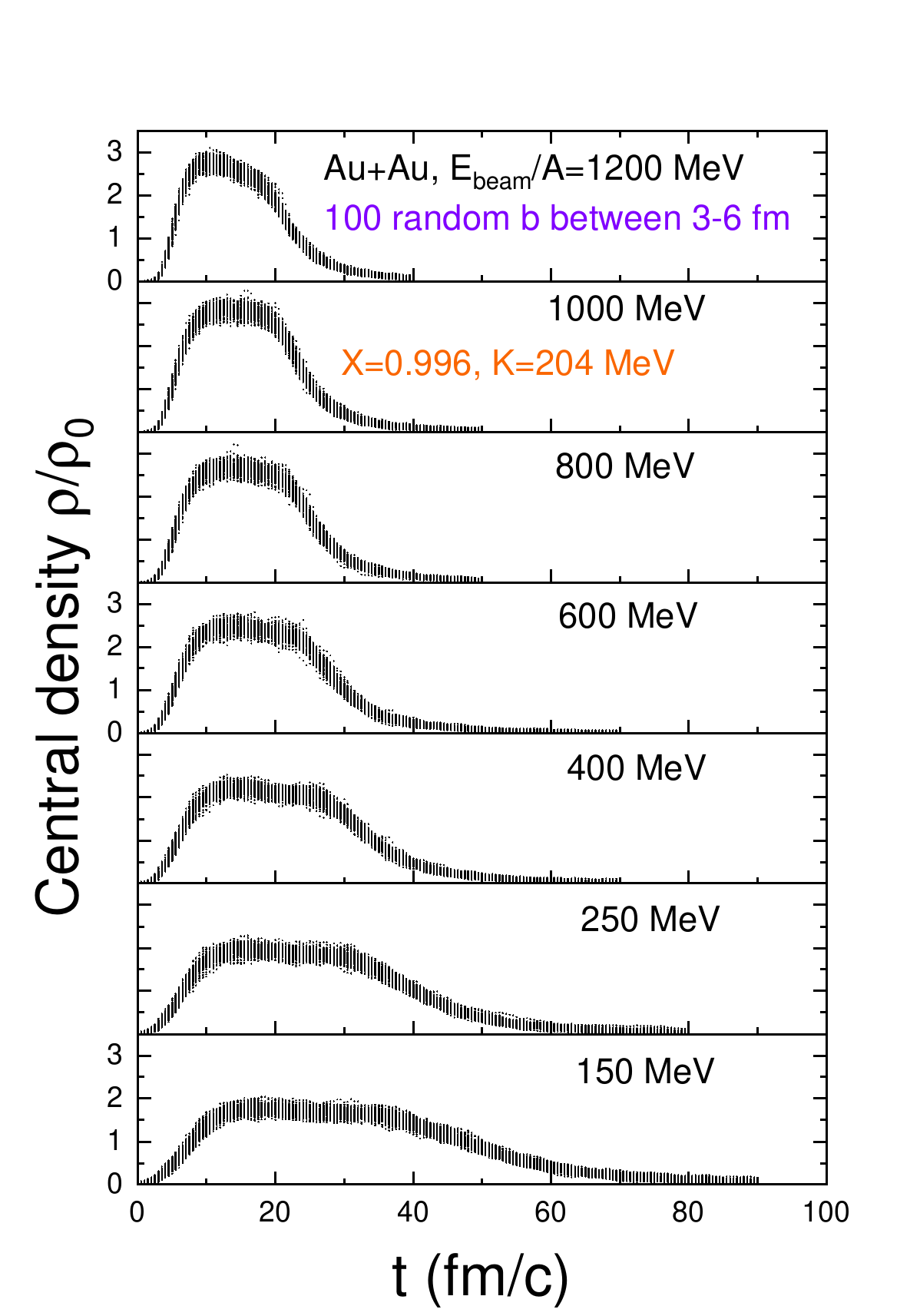}
}
\caption{Evolution of central baryon density during mid-central Au+Au reactions at beam energies between 150 to 1200 MeV/nucleon using a typically soft EOS ($K$=204 MeV) and approximately free-space BBSCSs ($X\approx 1$). At each instant, each dot represents the result of a specific impact parameter randomly generated between 3 and 6 fm.} \label{DT1}
\end{figure}
\begin{figure*}[thb]
\centering
 \resizebox{1.2\textwidth}{!}{
\includegraphics[width=3.5cm]{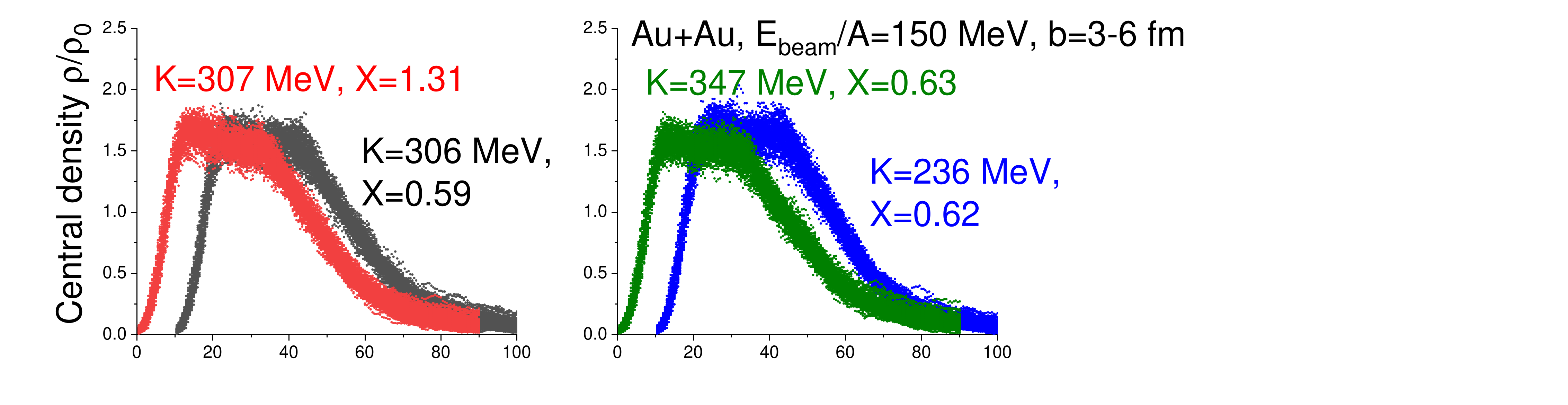}
}
 \resizebox{1.2\textwidth}{!}{
\includegraphics[width=3.5cm]{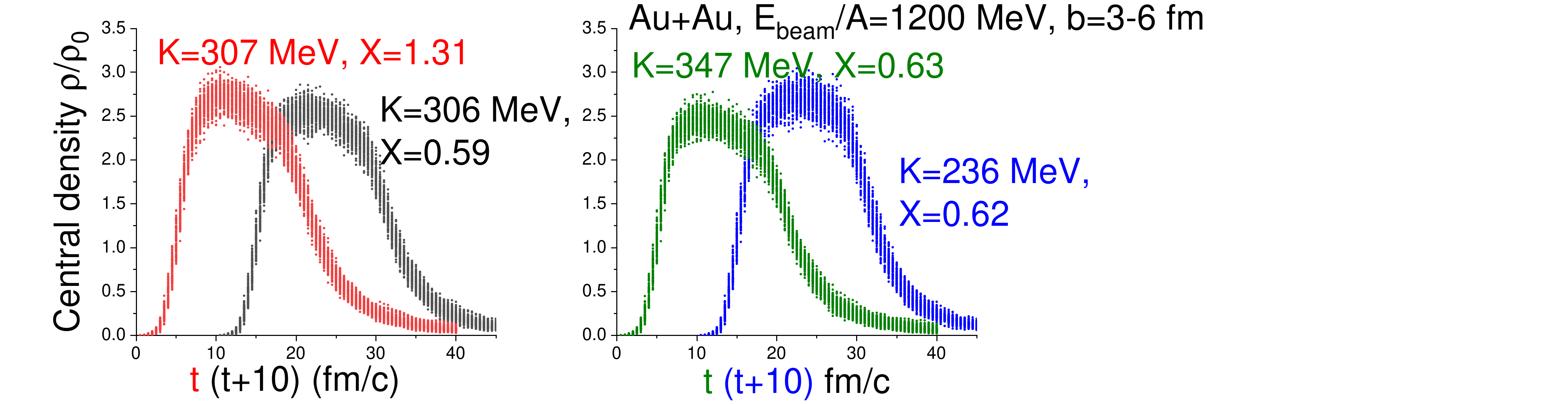}
}
\caption{Time evolutions of central baryon densities in mid-central Au+Au reactions at $E_{beam}/A$=150 MeV and 1200 MeV, respectively, using the four combinations of $X-K$ indicated. } \label{DT2}
\end{figure*}
\section{IBUU forward modeling of heavy-ion reactions}\label{Results}
In this section, we first report and discuss typical results of IBUU model predictions for compression and proton flow in heavy-ion reactions at intermediate energies. Using two examples, we illustrate respective effects of the $X$ and $K$ parameters on compression as well as proton directed and elliptical flow. 

\subsection{$X$ and $K$ effects on compression in mid-central Au+Au reactions from 150 to 1200 MeV/nucleon}\label{IBUU-P}
For studying the beam energy dependence of collective flow, it is useful to know firstly how the central density evolves. 
For this purpose, shown in Fig. \ref{DT1} are the evolution of central baryon density in a 1.0 fm$^3$ cubic cell around the c.m. of the two colliding nuclei) during mid-central Au+Au reactions using a typically soft EOS ($K$=204 MeV) and approximately free-space BBSCSs ($X\approx 1$). As we have carried out simulations for 120 sets of random combinations of $X$ and $K$ for each beam energy $E_{beam}/A$, we purposely selected this particular $X-K$ combination for an illustration. At each instant $t$, each one of the 100 dots represents the result of a specific impact parameter randomly generated between 3 and 6 fm. As expected, as the beam energy increases the maximum central density increases from about $1.7\rho_0$ at $E_{beam}/A$=150 MeV to about $2.7\rho_0$ at $E_{beam}/A$=1200 while its duration decreases gradually from about 30 to 15 fm/c. 

To evaluate individual effects of $X$ and $K$ on compression with respect to that due to the change of impact parameter within the centrality bin of 3-6 fm in the FOPI data used, selected from results using 120 sets of random $X-K$ combinations we compare in Fig. \ref{DT2}
the time evolutions of central densities in the Au+Au reactions at $E_{beam}/A$=150 MeV and 1200 MeV, respectively. In the two panels on the left, we selected the combinations of 
set-1 with $X=0.59$ and $K=306$ (black) while set-2 with $X=1.31$ and $K=307$ (red), respectively. They have approximately the same incompressibility (stiff) $K$ but their in-medium cross sections are different by about a factor of two. For a clear illustration, the x-axis (time t) of set-1 is shifted by 10 fm/c. It is seen that central density has little obvious difference at $E_{beam}/A$=150 MeV
in the two cases. While at $E_{beam}/A$=1200 MeV, doubling $X$ leads to a small but noticeable increase in compression. Similarly, shown in the two right panels are the results of using set-3 with $X=0.62$ and $K=236$ (blue) while set-4 with $X=0.63$ and $K=347$ (green), respectively. They have approximately the same $X$ but different incompressibilities corresponding to a typically soft EOS and stiff one, respectively. It is seen that at both beam energies the softer EOS leads to a higher central density as one expects. We notice that for the examples considered, the variations of the central density by varying the $X$ and $K$ values separately are compatible or smaller than the change due to the variation of the impact parameter between 3 and 6 fm (vertical width of the band). 

\subsection{$X$ and $K$ effects on proton collective flow }\label{XK-effect}
While the $X$ parameter has little (at $E_{beam}/A$=150 MeV) or small but noticeable (at $E_{beam}/A$=1200 MeV) effects on the evolution of the central density, as we shall show next, it does have significant effects on both directed and elliptical flows of free protons. First, lets recall some definitions and kinematics cuts used by the FOPI Collaboration in their data analyses. 
The first ($v_1$) and second ($v_2$) coefficients of the Fourier decomposition of particle azimuthal angle distribution $\frac{2\pi}{N}\frac{dN}{d\phi} = 1 + 2\sum_{n=1}^{\infty}v_n\cos{[n(\phi)]}$
are normally used to measure the strength of the so-called directed (transverse) and elliptical flow \cite{pawel85,oll,art}, respectively.
Their values (differential flows) at rapidity $y$ and transverse momentum $p_t$ can be evaluated from
$v_1(y,p_t)=\left<cos(\phi)\right>(y,p_t)=\frac{1}{n}\sum_{i=1}^{n}\frac{p_{ix}}{p_{it}}$ and
$v_2(y,p_{t})=\left<cos(2\phi)\right>(y,p_t)=\frac{1}{n}\sum_{i=1}^{n}\frac{p_{ix}^2-p_{iy}^2}{p_{it}^2}$, where $p_{ix}$ and $p_{iy}$ are the x- and y-component of the $i^{{\rm th}}$ particle momentum, respectively. In our setup of the simulations, the reaction plane is in the $x-o-z$ plane. Often, analyses are done in the center of mass (c.m.) frame of the two colliding nuclei and one uses nucleon's reduced rapidity $y'\equiv y/y_{\rm pro}$ where $y_{\rm pro}$ is the projectile rapidity in the cm frame. Moreover, the FOPI Collaboration used the reduced transverse
momentum $u_{t0}\equiv u_t/u_{pro}$ where
$u_{t} = \beta_{t} \gamma$ with respect to the speed of the projectile $u_{\text{pro}} = \beta_{\text{pro}} \gamma_{\text{pro}}$ where $\gamma = 1/\sqrt{1-\beta^{2}}$ \cite{FOPI}. We use the FOPI data obtained with the cut $u_{t0}\geq 0.8$ on transverse momentum. 
Moreover, the strength of directed flow 
is measured by using the slope  $v_{11}=F_1\equiv dv_1/dy'|_{y'=0}$ of proton directed flow $v_1$ at mid-rapidity. Similarly, the $v_2$ at mid-rapidity $v_{20}\equiv v_2(y'=0,u_{t0}\geq 0.8)$ is used to measure the strength of elliptical flow. 
\begin{figure*}[thb]
\centering
 \resizebox{0.499\textwidth}{!}{
\includegraphics[width=0.8\textwidth]{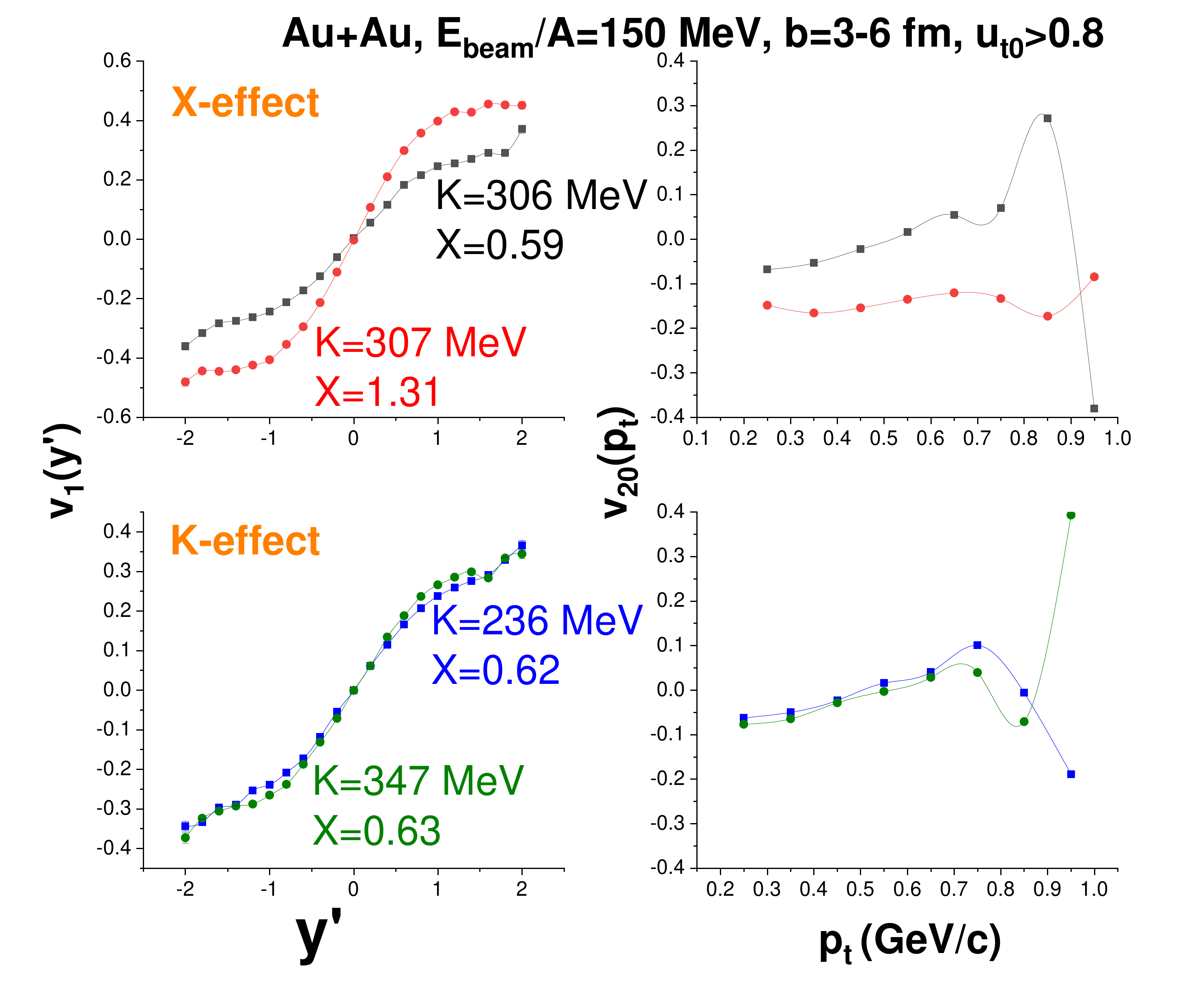}
}
 \resizebox{0.49\textwidth}{!}{
\includegraphics[width=0.8\textwidth]{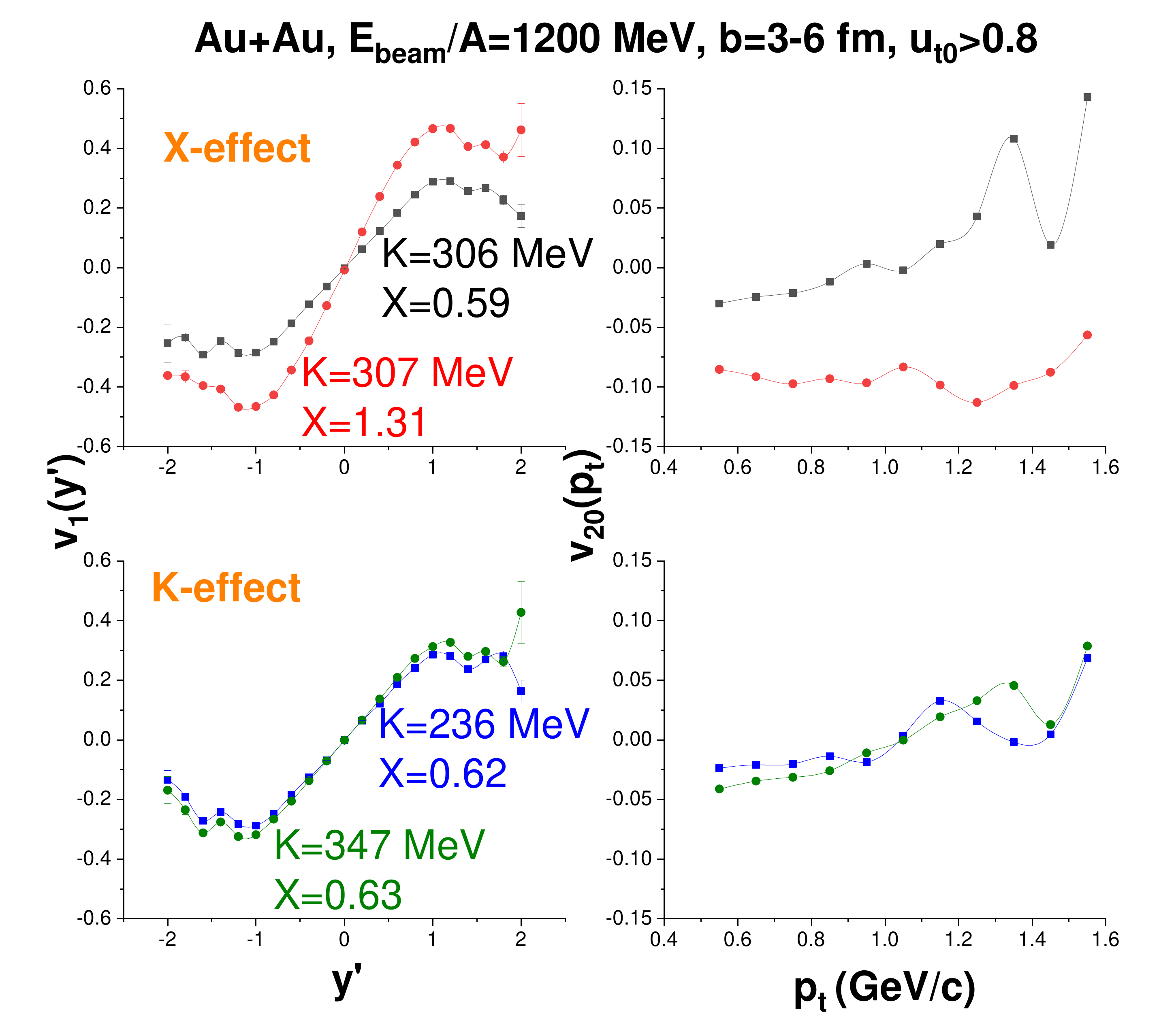}
}
\caption{Effects of $X$ (upper panels) and $K$ (lower panels) on the directed flow $v_1(y')$ and elliptical flow $v_{20}(p_t)$ at mid-rapidity for free protons with $u_{t0}\geq 0.8$ in mid-central Au+Au reactions at a beam energy of 150 and 1200 MeV/nucleon, respectively.}\label{v12E150}
\end{figure*}

In practice, in the FOPI data analyses their $v_1(y',u_{t0}\geq 0.8)$ data was fitted with an odd polynomial of $y'$ and the coefficient $v_{11}$ of its first term is taken as a measure of the strength of directed flow. It is equivalent to the slope $F_1$ of $v_1(y',u_{t0}\geq 0.8)$ with respect to rapidity at $y'=0$. For an easy automation necessary in our larger number of calculations, we approximated the slope at $y'=0$ with the average slope within $|y'| \leq 0.2$. As shown in Fig. \ref{v12E150}, the slope is not sensitive to the bin size used as long as $|y'| \leq 0.5$. Similarly, the FOPI Collaboration used an even polynomial of $y'$ to fit their $v_2(y',u_{t0}\geq 0.8)$ data and the constant term $v_{20}$ is taken to measure the strength of proton elliptic flow at mid-rapidity. We estimated it by taking the average of $v_2(y',u_{t0}\geq 0.8)$ within $|y'| \leq 0.2$. An inspection of the rapidity dependence $v_2(y',u_{t0}\geq 0.8)$ of the FOPI data in Ref. \cite{FOPI} indicates that such an estimation is accurate enough for the purpose of this work. It is also necessary to note that in all of our calculations, we identify free nucleons as those with local densities less than $\rho_0/8$ in the final state of the reaction. 

Fig. \ref{v12E150} demonstrates effects of $X$ (upper panels) and $K$ (lower panels) on the directed flow $v_1(y')$ and elliptical flow $v_{20}(p_t)$, from the IBUU calculations using the 4 sets of $X$ and $K$ parameters indicated in the plots and also used in Fig. \ref{DT2}. It is seen that both $X$ and $K$ affect significantly the directed and elliptical flow. Particularly, although the parameter $X$ has generally small effects on the evolution of central baryon density as discussed earlier, it affects significantly both directed and elliptical flows of protons. This is partially because it affects the temperature and density gradient 
obtained during heavy-ion reactions. 
While we have not studied thermalization in this work, an earlier work involving one of us \cite{LiChen05} using IBUU has shown clearly strong effects of X on thermalization and stopping power in heavy-ion reactions. 

It is evident that both $X$ and $K$ have opposite effects on $v_1(y')$ and $v_{20}(p_t)$ at mid-rapidity for protons with $u_{t0}\geq 0.8$. It is consistent with the feature of both the HADES data \cite{Ha2} and the FOPI data that the $v_{11}$ and $v_{20}$ are anti-correlated as shown in Fig. \ref{v12}. It is known that the positive transverse flow $v_{11}$ is developed earlier during the reaction due to the repulsive bounce-off between the spectators and participant region. While the elliptic flow $v_{20}$ is generated later, it varies between positive and negative in the beam energy range studied here. Because a larger $v_{11}$ created earlier in the reaction naturally reduces the spacial asymmetry in the transverse plane of the participant region where the $v_{20}$ is to be developed later, the transverse and elliptical flows are generally anti-correlated. 
It is also known that high $p_t$ particles dominate the $p_t$-integrated transverse flow $v_1(y')$ around both mid-rapidity and projectile/target rapidities \cite{Li-Jake}. Although there are only few high $p_t$ particles compared to mostly thermalized (isotropic) low $p_t$ particles, the large $<p_x>$ of these high $p_t$ particles dominate the transverse flow $v_1(y',p_t)$ especially around the interface of participant and spectator nucleons near the projectile/target rapidity. 
Selecting particles with transverse momenta higher than certain threshold, e.g., $u_{t0}\geq 0.8$ used by the FOPI Collaboration, will thus enhance the anti-correlation between the transverse and elliptical flows. This is because the total transverse momentum $p_t$ is limited by the cut $u_{t0}$ (this constraint is equivalent to a conservation law), generally speaking, if a larger $p_x$ has been developed for a particle at certain rapidity in the earlier stage of the reaction, there is less room to generate more $p_x$ later at the same rapidity. This will then further enhance the anti-correlation between the $v_{11}$ and $v_{20}$. 

We notice that with the 4 selected sets of $X$ and $K$ parameters, we can not make a fair comparison of their effects as they are not varied by the same percentage. As we shall show next, a more meaningful comparison of their relative effects will be done by using all 120 sets of $X$ and $K$ parameters varied in their entire prior uncertain ranges. It is also seen that at high transverse momentum, the $v_{20}(p_t)$ has large statistical fluctuations especially at $E_{beam}/A$=150 MeV as there are very few such high-momentum nucleons. Nevertheless, the FOPI data for $v_{20}\equiv v_2(y'=0,u_{t0}\geq 0.8)$ that we shall use in the Bayesian analysis is the integrated strength of elliptical flow from integrating $v_{20}(p_t)$ over all high momentum protons with $u_{t0}\geq 0.8$. Such quantity has reduced statistical fluctuations. 
\begin{figure*}[thb]
\centering
 \resizebox{0.49\textwidth}{!}{
\includegraphics[width=1\textwidth]{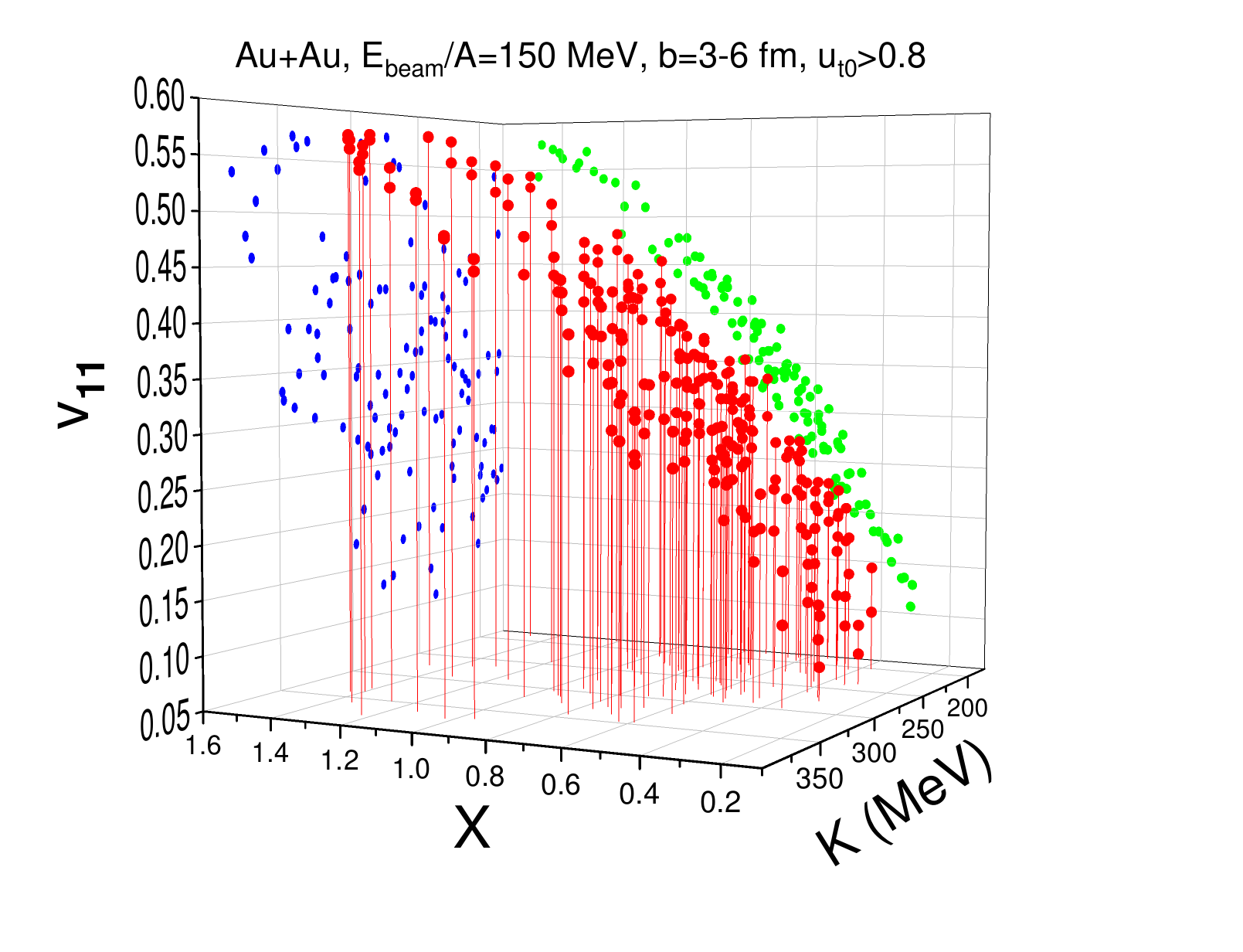}
}
 \resizebox{0.49\textwidth}{!}{
\includegraphics[width=1\textwidth]{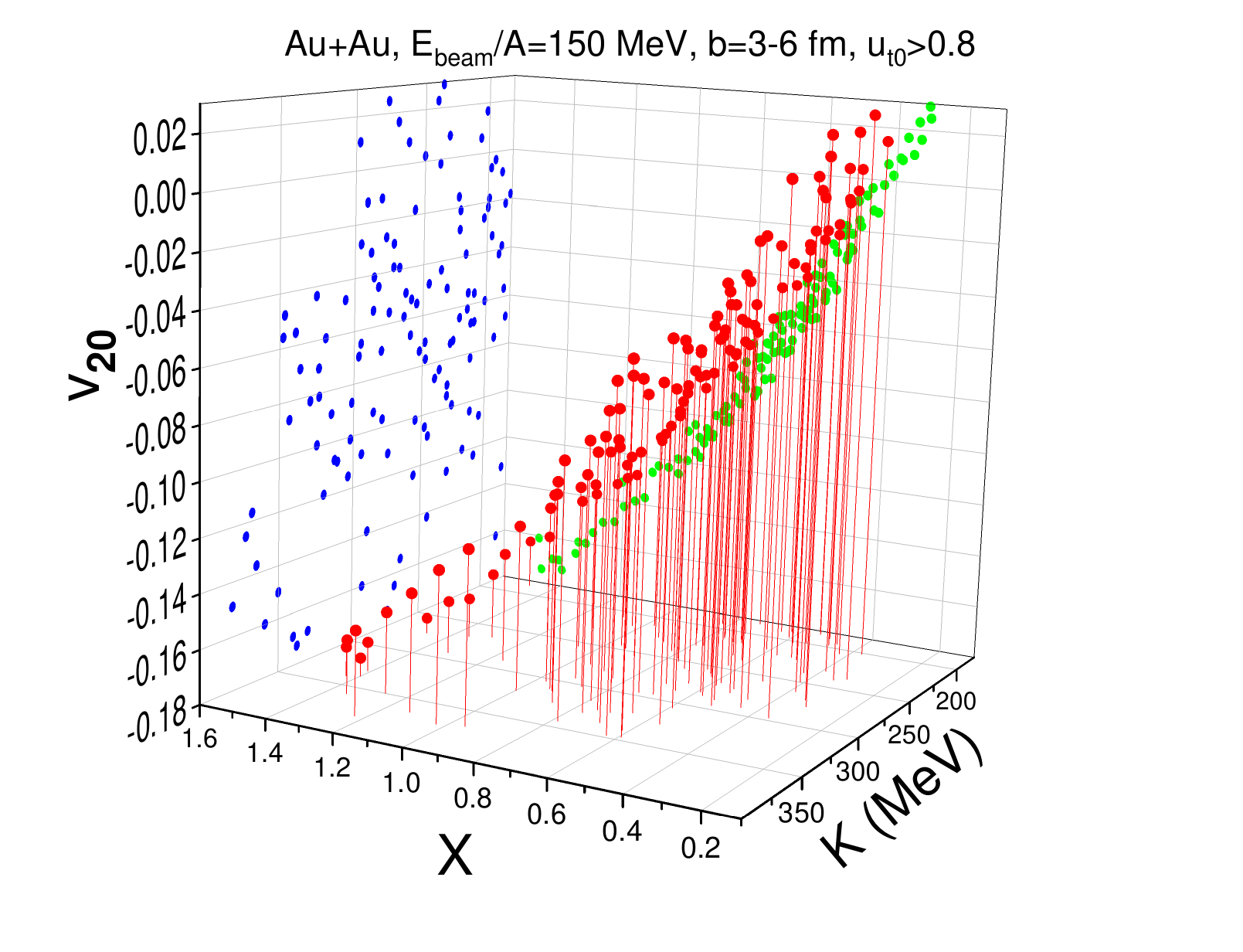}
}
\caption{Left: IBUU predictions for the proton directed flow strength $v_{11}=F_1$ with 120 sets of X and K parameters for the mid-central Au+Au reactions at a beam energy of 150 MeV/nucleon. Right: The strength of proton elliptical flow $v_{20}$ at mid-rapidity from the same sets of IBUU calculations. The blue and green dots are projections to the respective 2-dimensional planes.   
}\label{train1}
\end{figure*}

\begin{figure*}[thb]
\centering
 \resizebox{0.49\textwidth}{!}{
\includegraphics[width=1\textwidth]{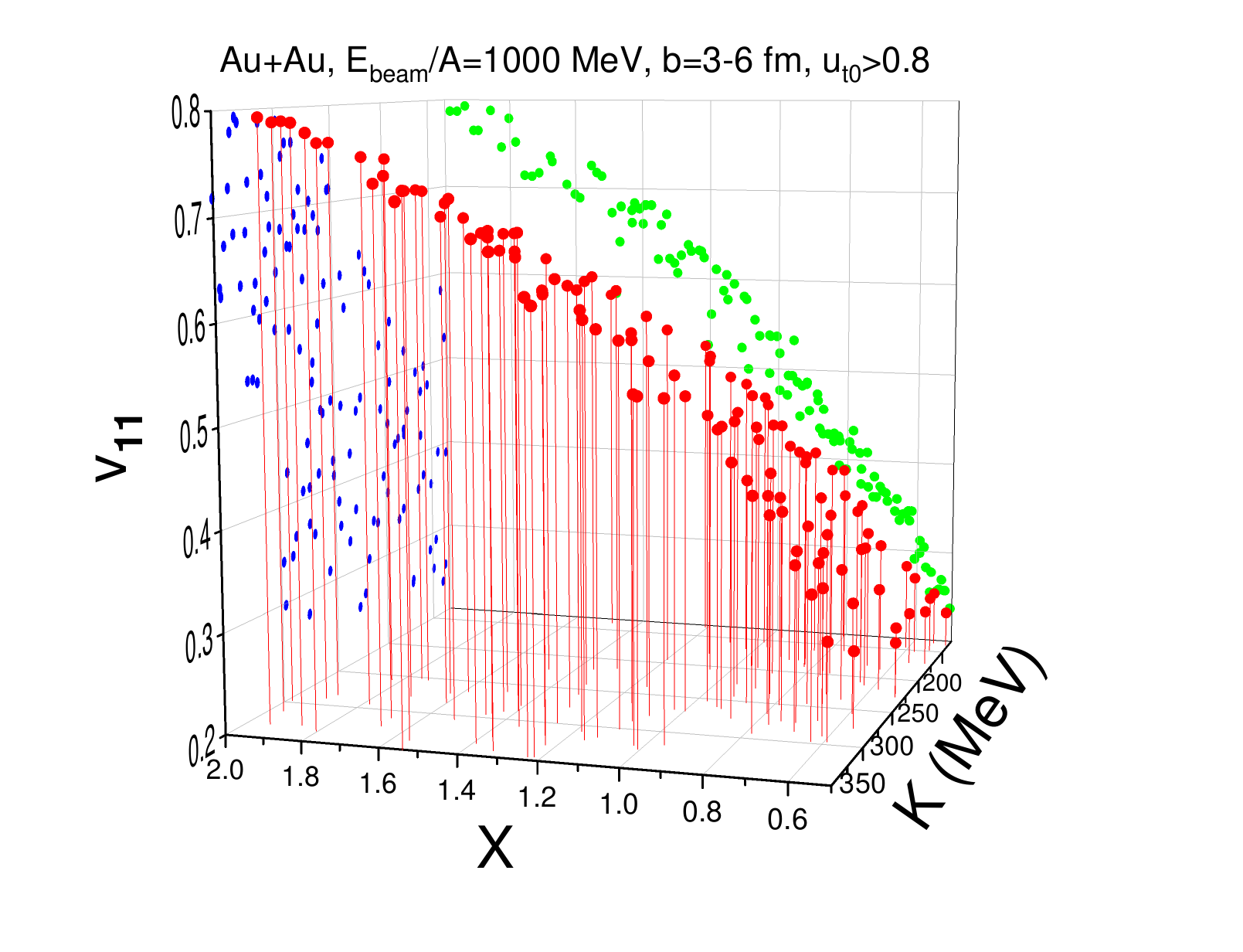}
}
 \resizebox{0.49\textwidth}{!}{
\includegraphics[width=1\textwidth]{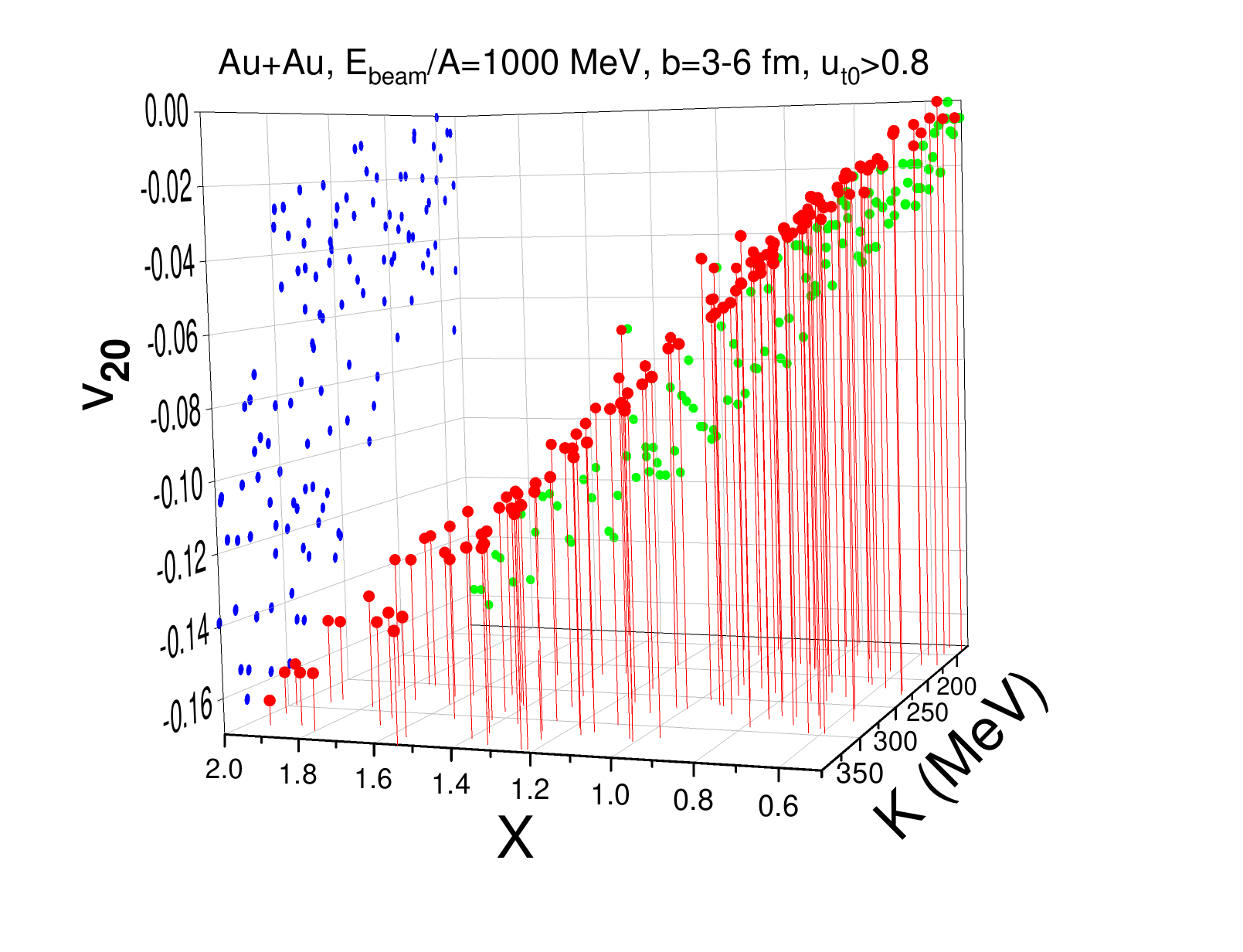}
}
\caption{The same as in Fig. \ref{train1} but at a beam energy of 1000 MeV/nucleon. }\label{train2}
\end{figure*}

\subsection{Training data for the GP emulator of IBUU}\label{training}

We performed IBUU simulations with 120 sets of $X$ and $K$ parameters generated randomly on the Latin hyperlattice within their prior ranges discussed earlier. As two examples, shown in Fig. \ref{train1} and Fig. \ref{train2} are 3-dimensional scatter plots of $v_{11}$ and $v_{20}$ as functions of $X$ and $K$ with $E_{beam}/A=150$ MeV and 1000 MeV, respectively. To see more clearly individual effects of $X$ and $K$, the results are projected to the respective 2-dimensional planes as shown with the blue and green dots, respectively. In all cases we studied, both $v_{11}$ and $v_{20}$ are more strongly correlated with $X$ than $K$ as indicated by the rather narrow bands of green dots and the pretty widely scattered blue dots. It is also seen that the $v_{20}$ is more strongly correlated with $K$ than $v_{11}$.  
Consequently, for a given pair of $v_{11}$ and $v_{20}$ data within their error bars, an approximately unique value of $X$ can be determined with some quantified uncertainty. However, multiple $K$ values may exist to give approximately the same $v_{11}$ and $v_{20}$ values especially at lower beam energies. Thus, even if the in-medium BBSCSs are well determined, it is still very challenging to determine the stiffness parameter $K$ using the observables considered here. 

Considering all $v_{11}$ and $v_{20}$ data available in heavy-ion reactions at different beam energies, to our best knowledge, there is no physics reason to expect a constant $K$ as the stiffness of hot and dense matter formed is changing with the conditions of these reactions. In our opinion, this result is not an indication of any drawback of the Skyrme energy density function in which the single parameter $K$ describes the stiffness of cold nuclear matter in the entire density range. We emphasize that reactions at different beam energies probe stiffness of hot and dense matter at different densities and temperatures. The parameter $K$ inferred from relatively low-energy heavy-ion reactions reflects effects of the incompressibility around $\rho_0$, while its values obtained from relatively high-energy heavy-ion reactions reflect the high-density stiffness of cold dense matter characterized by the skewness $J_0$ and kurtosis $Z_0$ parameters as shown in Fig. \ref{J0Z0K}. Thus, it is only meaningful to compare the parameter $K$ from relatively low-energy heavy-ion reactions with the incompressibility $K_0$ from collective oscillation modes (e.g., giant resonances) or structures of heavy nuclei. Of course, in a relatively high-energy high-energy collisions, the varying stiffness of nuclear matter in the whole density region may play important roles. It is thus useful to investigate how the stiffness of nuclear matter may evolve with density and temperature by analyzing the flow excitation functions. In this regard, the FOPI data is invaluable.

Going beyond the Skyrme energy density functional with or without considering the momentum-dependence of single particle potentials, it is possible to vary independently the EOS parameters relevant at saturation (e.g., $K_0$) and supra-saturation (e.g., $J_0$ and $Z_0$) densities, respectively. This can be done by using, e.g., a meta-model EOS \cite{NBZ18,Li24} by parameterizing the $E_0(\rho)$ using at least three independent variables ($K_0$, $J_0$ and $Z_0$), or piecewise polytropes for several connected density regions \cite{LBM}, similar to many studies on neutron stars, see, e.g., Ref. \cite{LiBA21} for a review. While these approaches allow constraints on the presumably independent parameters characterizing explicitly the stiffness of nuclear matter at different densities, they have their own challenges and limitations. For example, in the case of using a meta-model EOS, we have carried out a Bayesian analysis \cite{Xie20} of the pressure of cold nuclear matter in the density range of about $1.3\rho_0$ to $4.5\rho_0$ from transport model analyses of kaon production and nuclear collective flow in relativistic heavy-ion collisions \cite{Pawel02,lynch09}. It was found that the posterior PDFs of $J_0$ and $Z_0$ depend strongly on the prior range and PDF of $K_0$ adopted. In the case of using the piecewise polytropes, to our best knowledge, no analysis has been done for heavy-ion reactions. However, it is known in the literature from studying neutron stars that there is little or no direct information in setting the prior range and PDF of parameters describing the high-density segments of the polytropic EOS. In fact, they sequentially depend on the prior range and PDF of the EOS parameters around $\rho_0$. These analyses also often require a lot more than two parameters, thus training and validating emulators in larger multi-dimensional parameter space as well as significantly more computing time for Bayesian analyses are necessary. 

We are making progress in using a meta-model EOS in analyzing the same heavy-ion reaction data, and the results will be reported elsewhere. Besides having their own important scientific merits, results of our work reported here using the Skyrme energy density functional provide a useful reference or guidance for future studies with other EOSs. 

\begin{figure}[thb]
\vspace{-1cm}
\centering
 \resizebox{0.55\textwidth}{!}{
\includegraphics[width=1\textwidth]{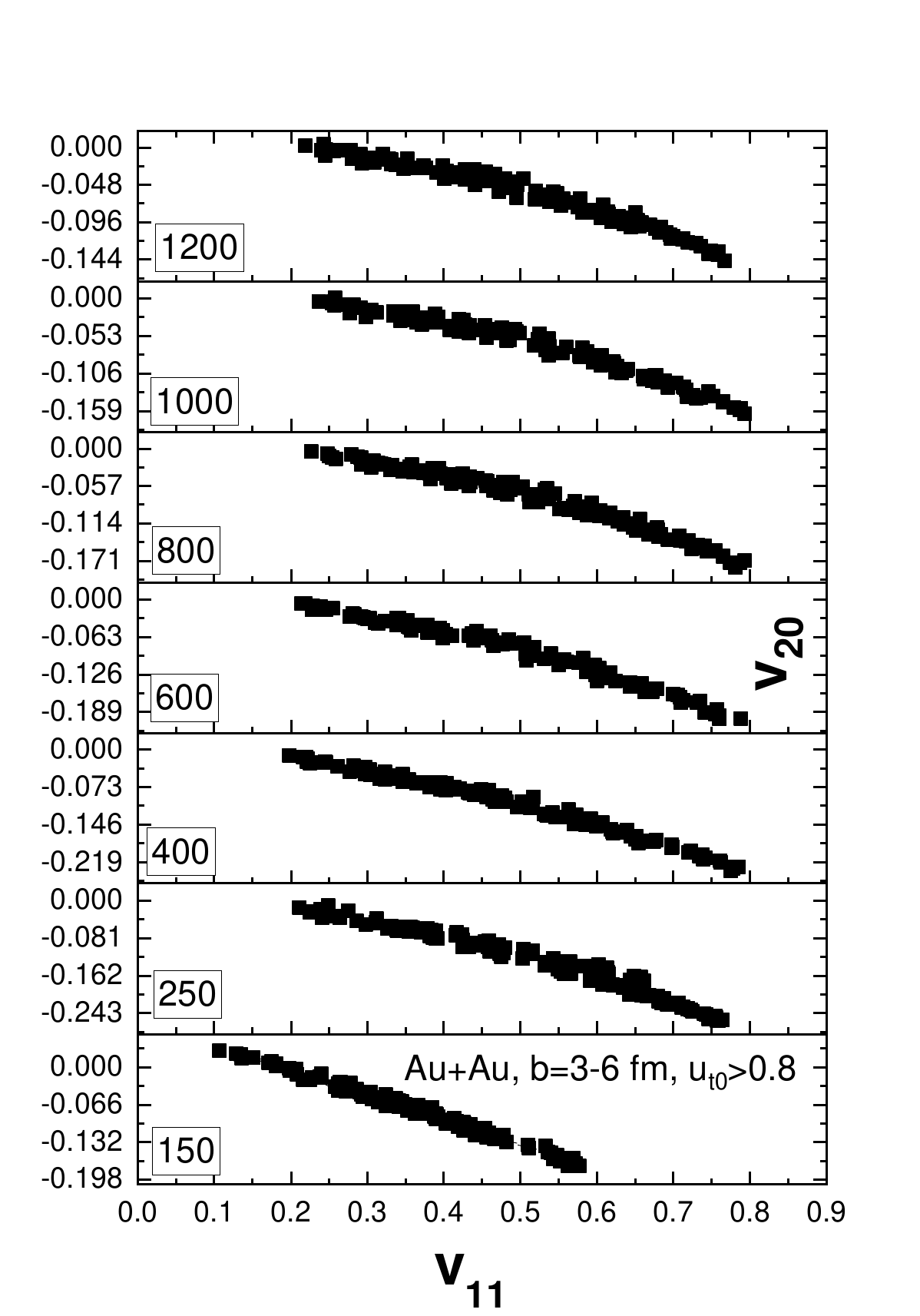}
}
\caption{Anti-correlation between $v_{20}$ and $v_{11}=F_1$ for free protons in mid-central Au+Au collisions at beam energies from 150 to 1200 MeV/nucleon.} \label{v1v2-EC}
\end{figure}
\begin{figure}[thb]
\centering
 \resizebox{0.5\textwidth}{!}{
\includegraphics[width=1\textwidth]{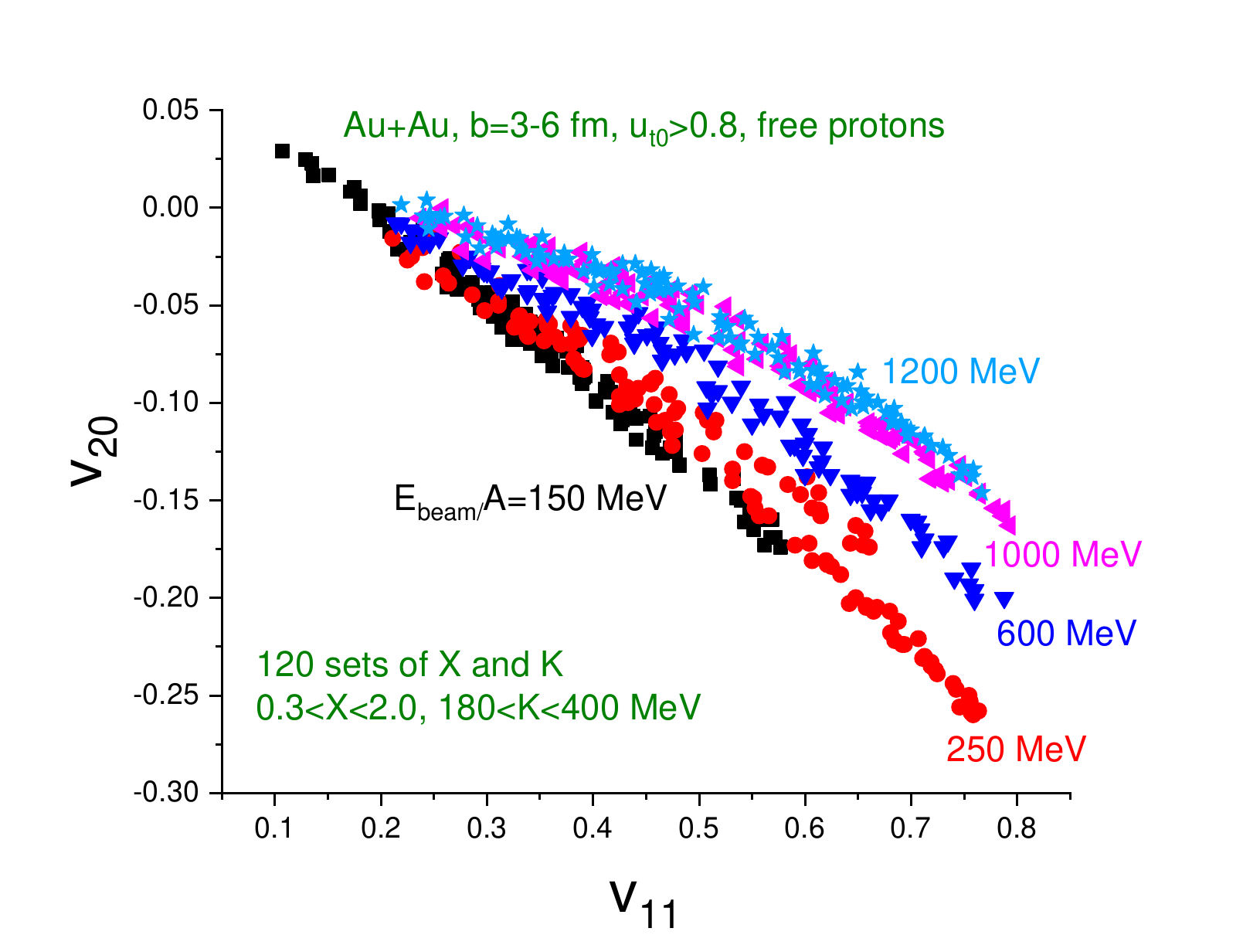}
}
\caption{Anti-correlation between $v_{20}$ and $v_{11}=F_1$ for protons at the selected beam energies.} \label{c12}
\end{figure}

\subsection{Anti-correlations between directed and elliptical flow of free protons}\label{acr}
To evaluate more quantitatively the (anti-)correlations of $v_{11}$ and $v_{20}$ as well as their maximum values reachable by varying $X$ and $K$ within their prior uncertainty ranges, shown in Fig. \ref{v1v2-EC} is a  compilation of $v_{20}$ versus $v_{11}$
at beam energies from 150 to 1200 MeV/nucleon. Generally, as the $v_{11}$ increases, the $v_{20}$ decrease gradually, reflecting a strong anti-correlation between them. It is also seen that the maximum value of $v_{11}$ first increases with beam energy then saturates at around 0.8 when $E_{beam}$ becomes higher than about 400 MeV/nucleon. On the other hand, the magnitude (absolute value) of $v_{20}$ shows a stronger dependence on the beam energy. Its maximum magnitude first increases to about 0.25 at a beam energy of about 250 MeV/nucleon, then decreases with increasing beam energy. These features are consistent with those of the FOPI data shown in Fig. \ref{v12}.

A more clear and quantitative comparison of $v_{11}$ versus $v_{20}$ on the same scale is given in Fig. \ref{c12}. To avoid overlapping, only results of 5 beam energies are shown. As shown in Fig. \ref{train1}, at 150 MeV/nucleon beam energy, the $v_{20}$ can be as high as +0.03 obtained with small $X$ and $K$ values. On the other hand, at the same beam energy of 150 MeV/nucleon with large $X$ and $K$ values, $v_{20}$ can be as low as $-0.18$. While for $v_{11}$, its maximum value increases from about 0.6 at 150 MeV/nucleon to about 0.8 at 400 MeV/nucleon, then saturates. At beam energies above about 250 MeV/nucleon, the $v_{20}$ is always negative and its correlation with $v_{11}$ becomes weaker. Above about 800 MeV/nucleon beam energy, the results with different beam energies largely overlap, indicating a rather weak beam energy dependence of both $v_{20}$ and $v_{11}$ with these high beam energies. 

\subsection{Validations of GP emulations of IBUU prediction of proton flows}
\begin{figure*}
\centering
\includegraphics[width=0.47\textwidth,height=0.65\textwidth]{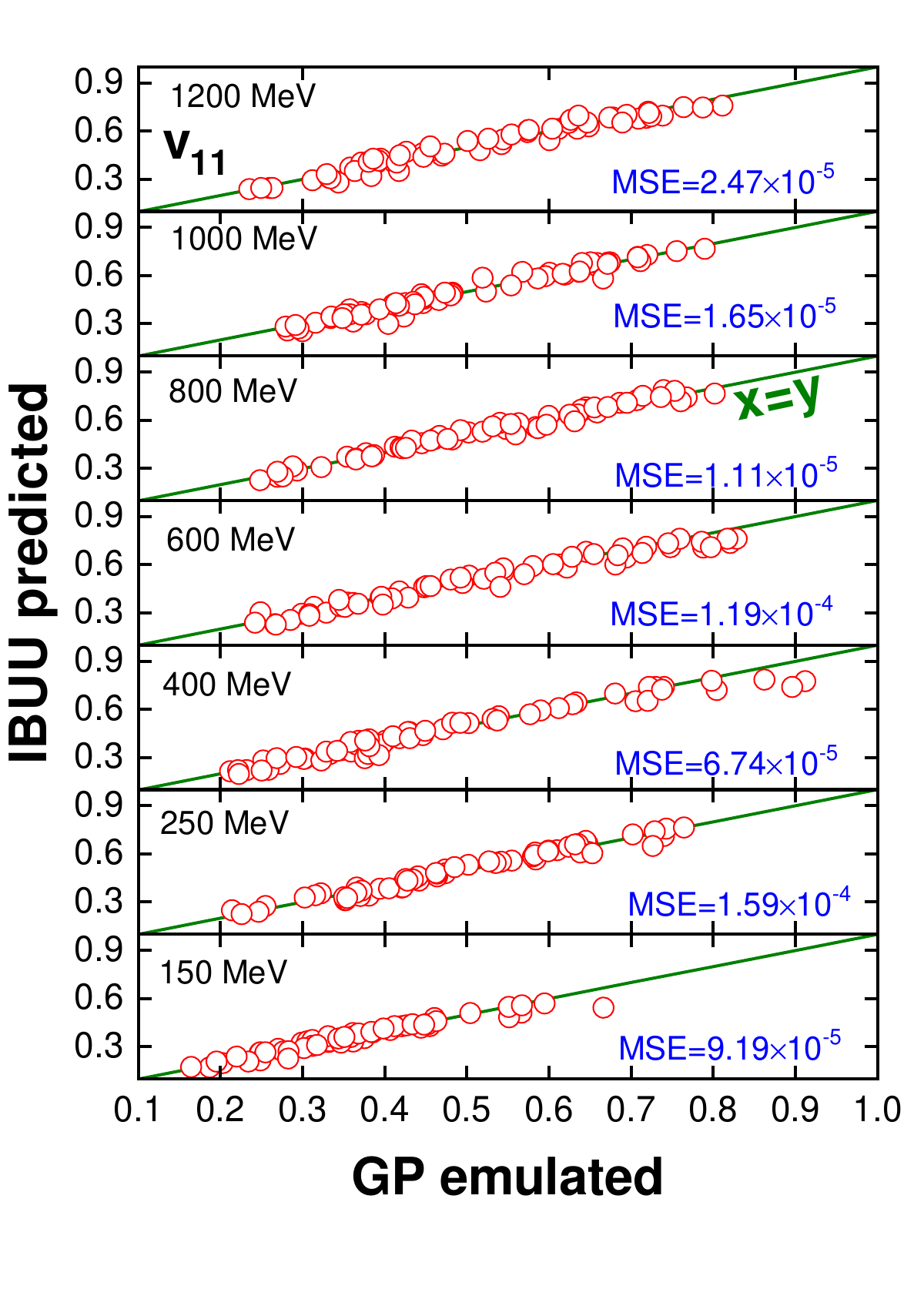}
\includegraphics[width=0.49\textwidth,height=0.65\textwidth]{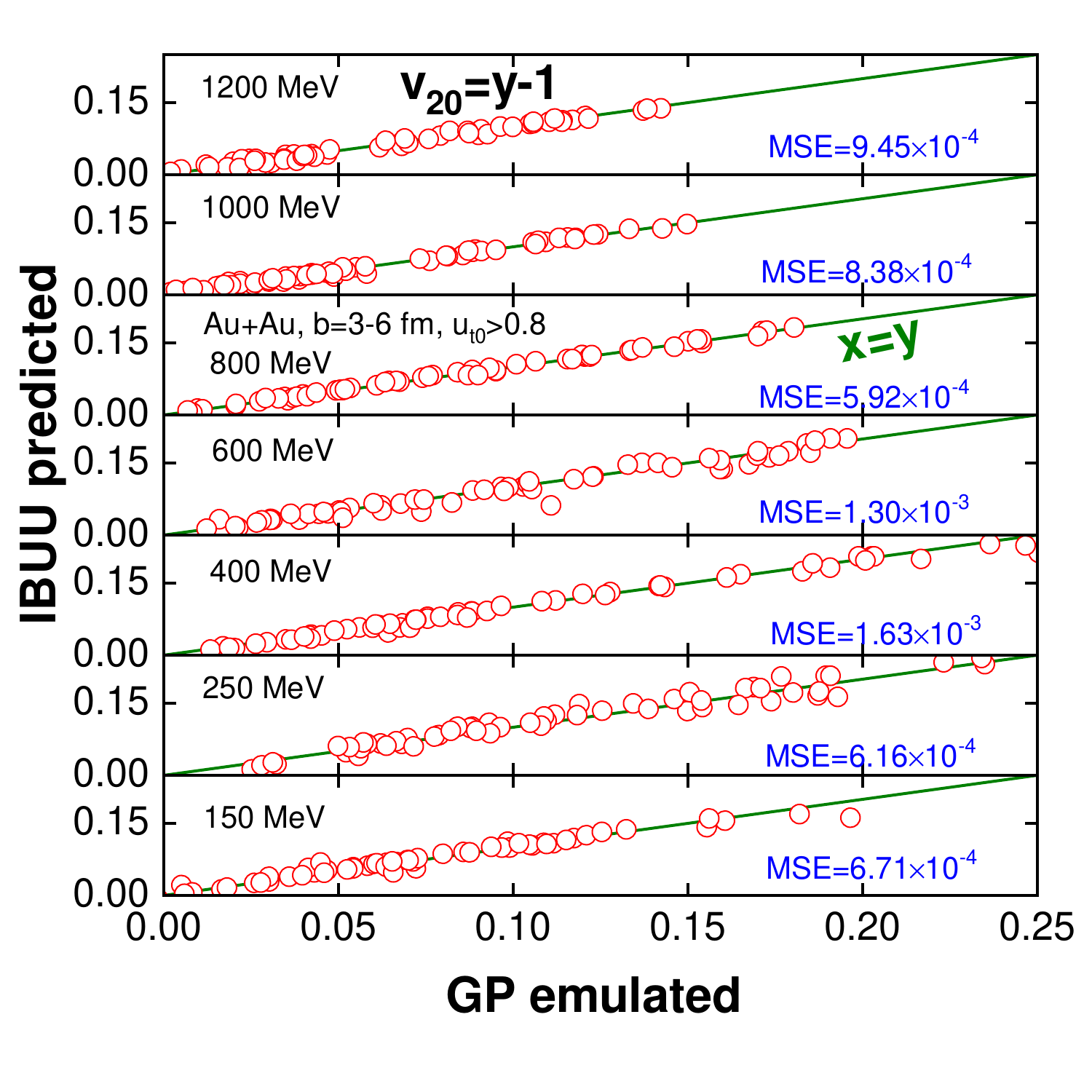}
\caption{Validation of GP emulations of IBUU predictions for $v_{11}$ (left) and $v_{20}$ (right) for the mid-central Au+Au reactions at beam energies from 150 to 1200 MeV/nucleon. The GP emulations and IBUU predictions would be equal if they fall on the diagonal $x=y$ lines (green). Because $v_{20}$ is normally very small and negative, to evaluate the quality of emulating $v_{11}$ and $v_{20}$ on the same scale, shown in the right panels are the values of $1+v_{20}$ (namely $v_{20}=y-1$ as indicated).
} \label{GP}
\end{figure*}
One method to validate the GP emulator is to directly compare the emulation results with simulator predictions with model parameters that are not used in the training of the emulator. The degree of agreement can be measured quantitatively by using the mean squared errors (MSE) or visually by plotting the results of simulations versus those from emulations. As in our previous work \cite{LX-HA} in Bayesian analyses of the HADES data, we have carried out such tests at each beam energy in analyzing here the FOPI data. As examples, shown in Fig. \ref{GP} 
are comparisons of GP emulations of IBUU predictions for $v_{11}$ (left) and $v_{20}$ (it is equal to the plotted value-1 for ease of presentation) (right). The GP emulations and IBUU predictions would be equal if they fall on the $y=x$ lines (green). Generally speaking, the trained GP emulators can reproduce the independent IBUU predictions very nicely for all beam energies considered. 

To be more quantitative, we also indicated the MSE values in the plots. The MSE between the GP output $O_{\rm{GP,i}}$ and the IBUU prediction $O_{\rm{IBUU,i}}$ for the observable $O=v_{11}~\rm{or}~v_{20}$ is defined as
\begin{equation}
    {\rm MSE} = \frac{1}{n} \sum^{n}_{i = 1} (O_{\rm{GP,i}} - O_{\rm{IBUU,i}})^2
\end{equation}
where n=60 is the number of tests made for this presentation using another 60 IBUU calculations as the training data set. It is seen that the MSE values for both $v_{11}$ and $v_{20}$ are all very small. We notice  that the MSE for $v_{20}$ is about 10 times larger than that for $v_{11}$, indicating the intrinsic difficulties in emulating the generally small second-order proton flow. Nevertheless, comparing the square roots of MSE values with the mean values of $v_{11}$ and $v_{20}$ emulated in the whole ranges of $X$ and $K$ parameters, the relative accuracies of the GP emulations are all reasonably good. 

In this example of testing, a 50/50 split of the 120 sets of IBUU predictions is used for training/validating the emulators. The overall rather good agreements between results from the GP emulator and IBUU simulator give us the confidence that our GP emulators are accurate. Depending on how quickly and easily one can generate training/validating data, there is really no hard and fast rule on the optimal training/validating split ratio \cite{Ra,Nick}. To take complete advantage of our time-consuming IBUU calculations, in our final Bayesian analyses after extensive tests we use all 120 sets of IBUU calculations at each beam energy in training our GP emulators. Thus, the actual accuracies of GP emulators used in our final analyses are expected to be better than those shown in Fig. \ref{GP}.

\begin{figure*}[thb]
 \resizebox{0.92\textwidth}{!}{
\includegraphics[width=1\textwidth]{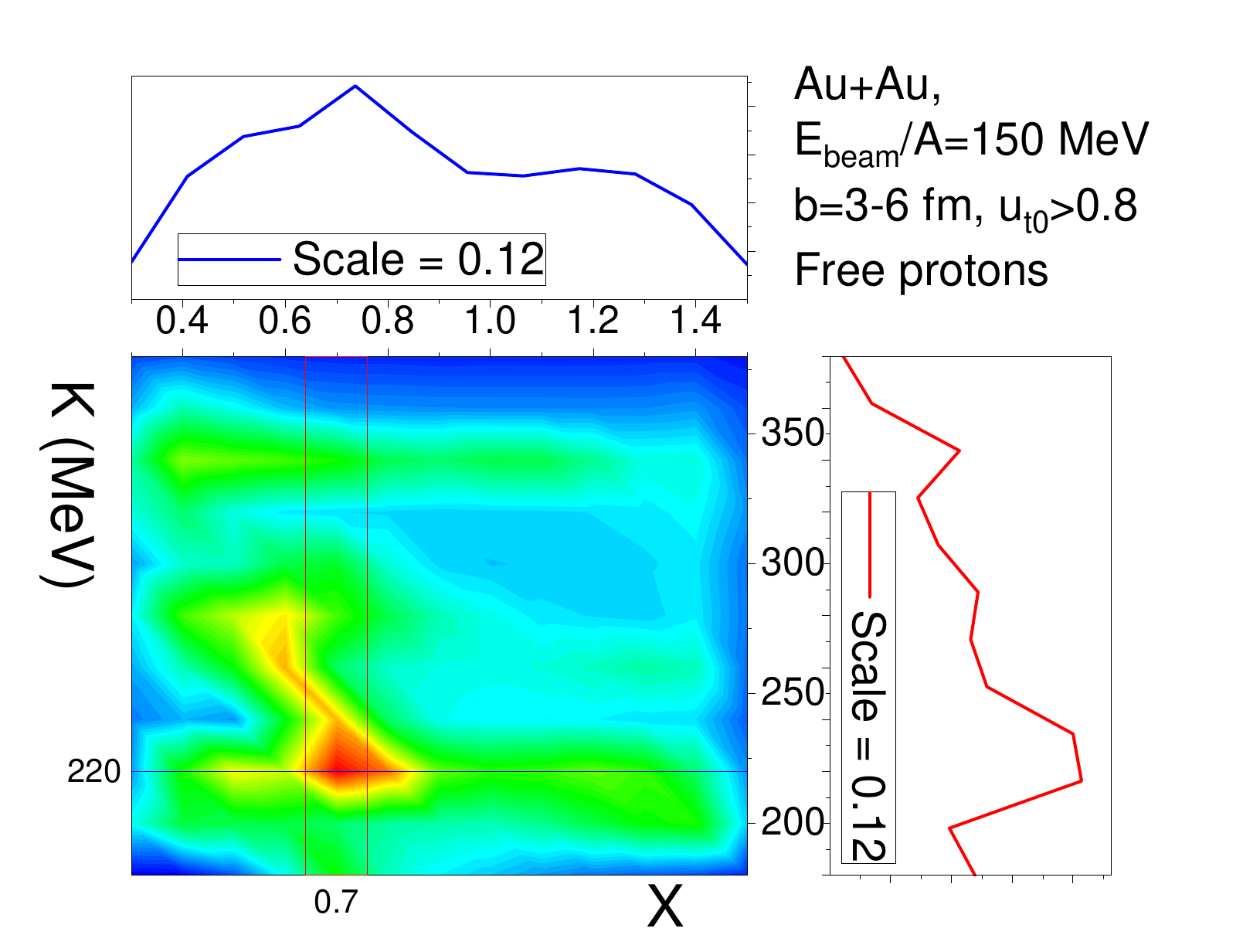}
}
\hspace{2cm}
 \resizebox{1.4\textwidth}{!}{
\includegraphics[width=1.2\textwidth]{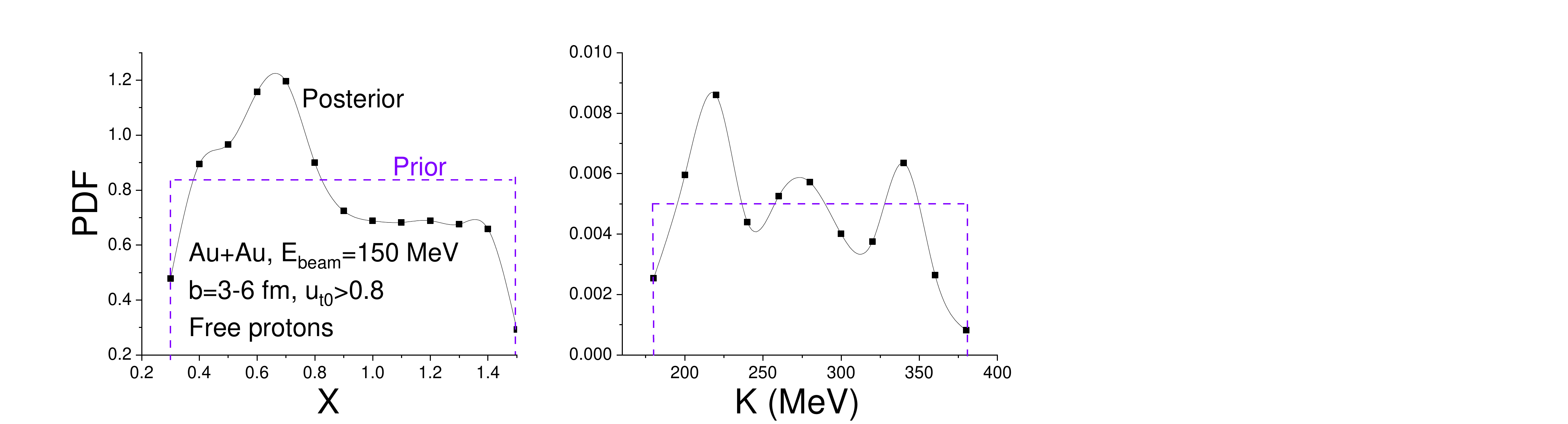}
}
\caption{Upper panels: Posterior correlations (heat map) of $X$ and $K$ and their profiles (on the upper and right sides) along the lines indicated. Lower panels: the marginalized posterior PDFs (solid) of $X$ and $K$, respectively, with respect to their prior PDFs (dashed rectangles). These results are inferred from the FOPI  proton $v_{20}$ and $v_{11}$ data in mid-central Au+Au reactions at a beam energy of 150 MeV/nucleon.}\label{ProfileE150}
\end{figure*}

\begin{figure*}[thb]
\centering
 \resizebox{0.92\textwidth}{!}{
\includegraphics[width=1.\textwidth]{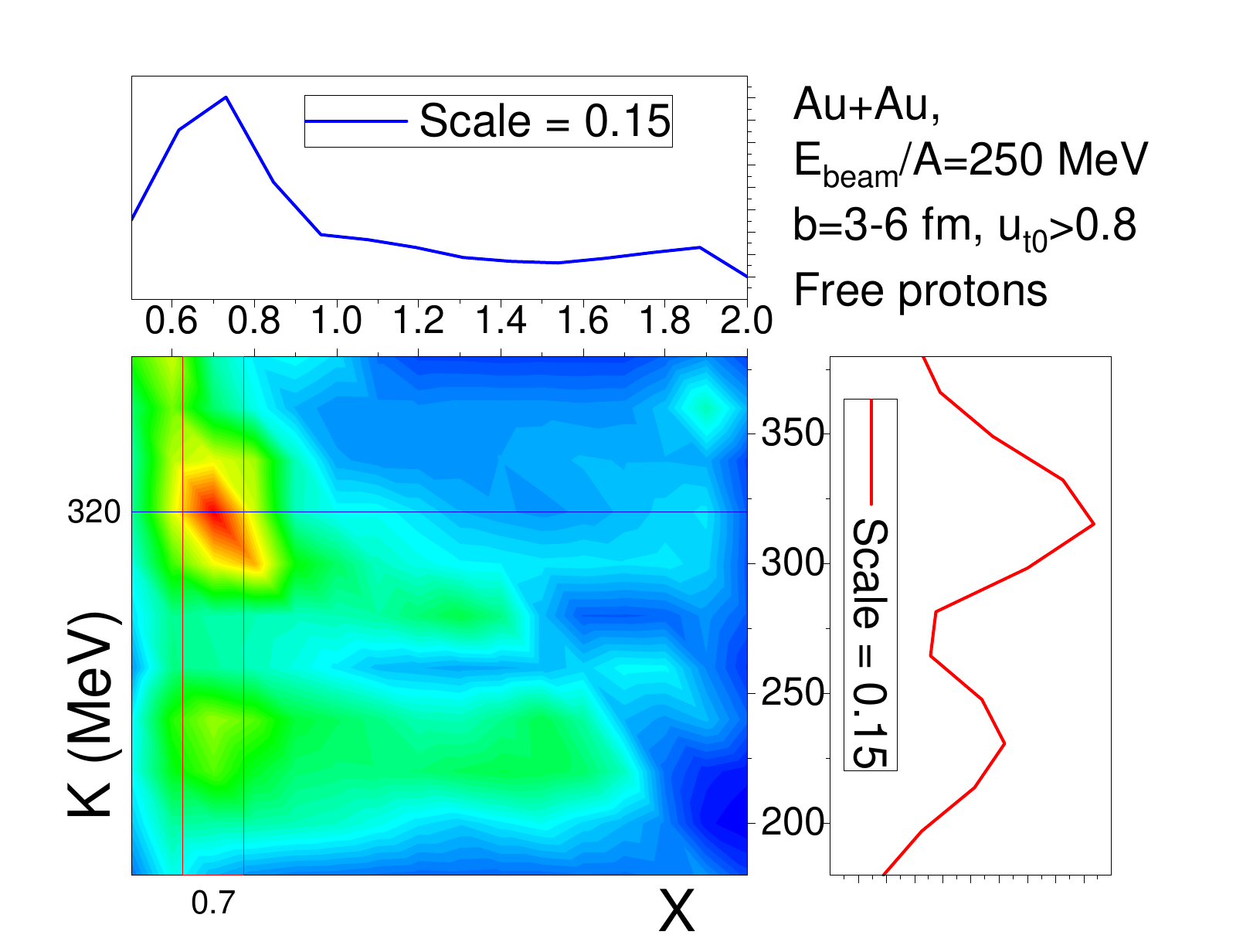}
}
 \resizebox{1.4\textwidth}{!}{
\includegraphics[width=1.2\textwidth]{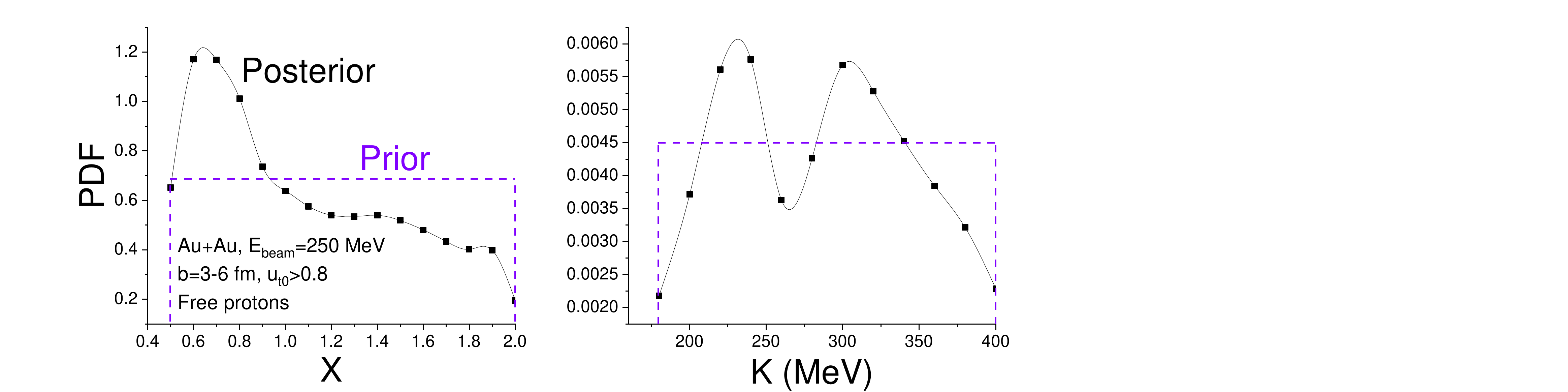}
}
\caption{Same as in Fig. \ref{ProfileE150} but using the FOPI  proton $v_{20}$ and $v_{11}$ data in mid-central Au+Au reactions at a beam energy of 250 MeV/nucleon.}\label{ProfileE250}
\end{figure*}

\begin{figure*}[thb]
\centering
 \resizebox{0.92\textwidth}{!}{
\includegraphics[width=1.\textwidth]{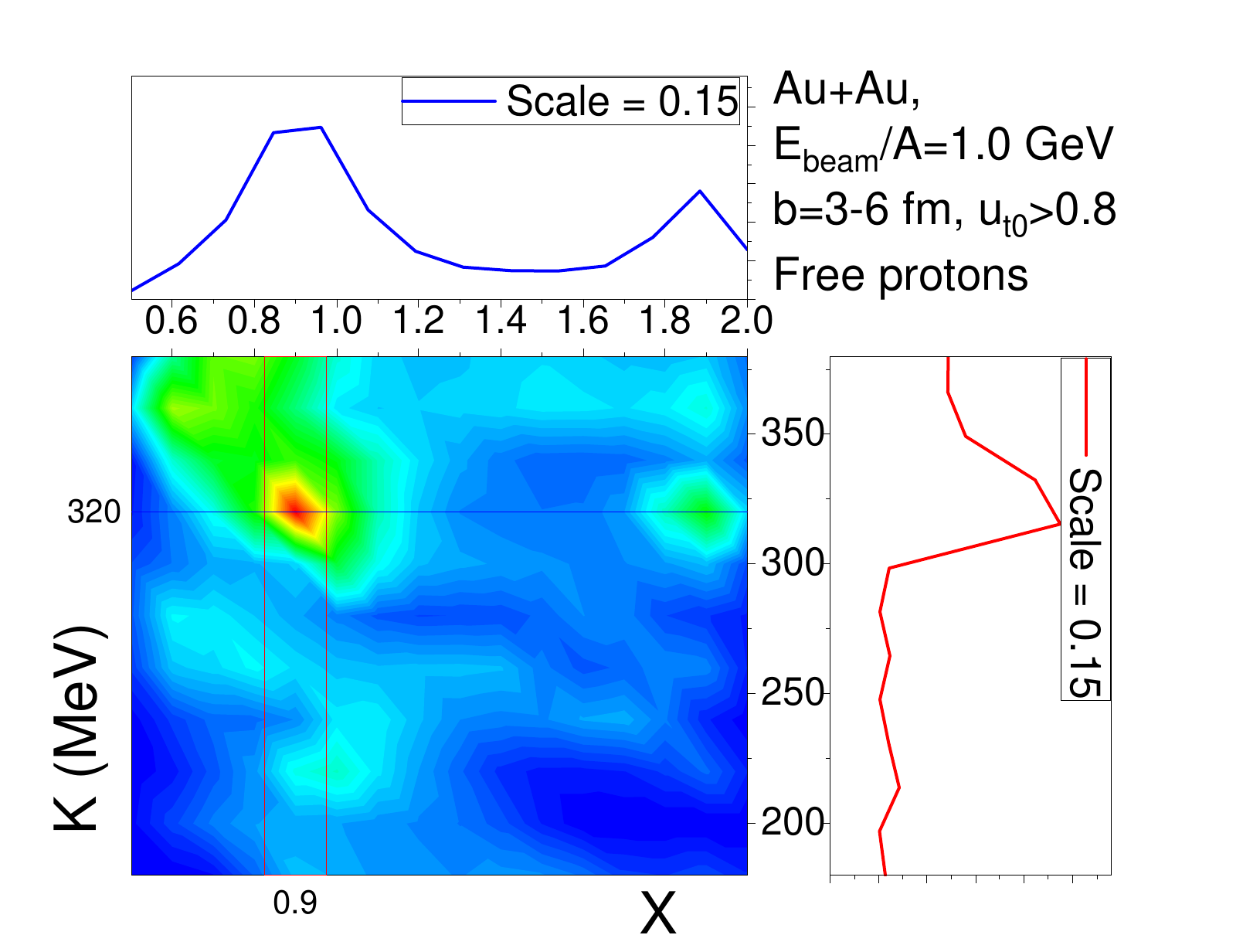}
}
 \resizebox{1.4\textwidth}{!}{
\includegraphics[width=1.2\textwidth]{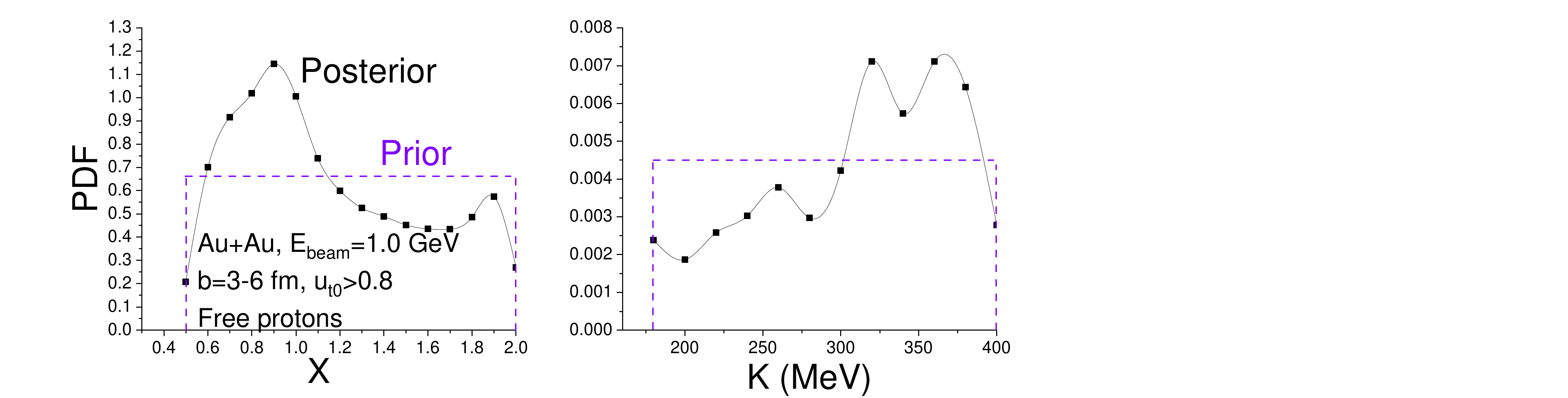}
}
\caption{Same as in Fig. \ref{ProfileE150} but using the FOPI  proton $v_{20}$ and $v_{11}$ data in mid-central Au+Au reactions at a beam energy of 1000 MeV/nucleon.}\label{ProfileE1000}
\end{figure*}

\section{Bayesian inference of posterior PDFs of $X$ and $K$ as well as their correlations and beam energy dependence}\label{B-results}
Here we present and discuss results of our Bayesian analyses of the FOPI proton flow excitation functions shown in Fig. \ref{v12}. We first give three typical examples of posterior PDFs before presenting their beam energy dependence.
\begin{figure*}[thb]
\centering
 \resizebox{0.49\textwidth}{!}{
\includegraphics[width=1\textwidth]{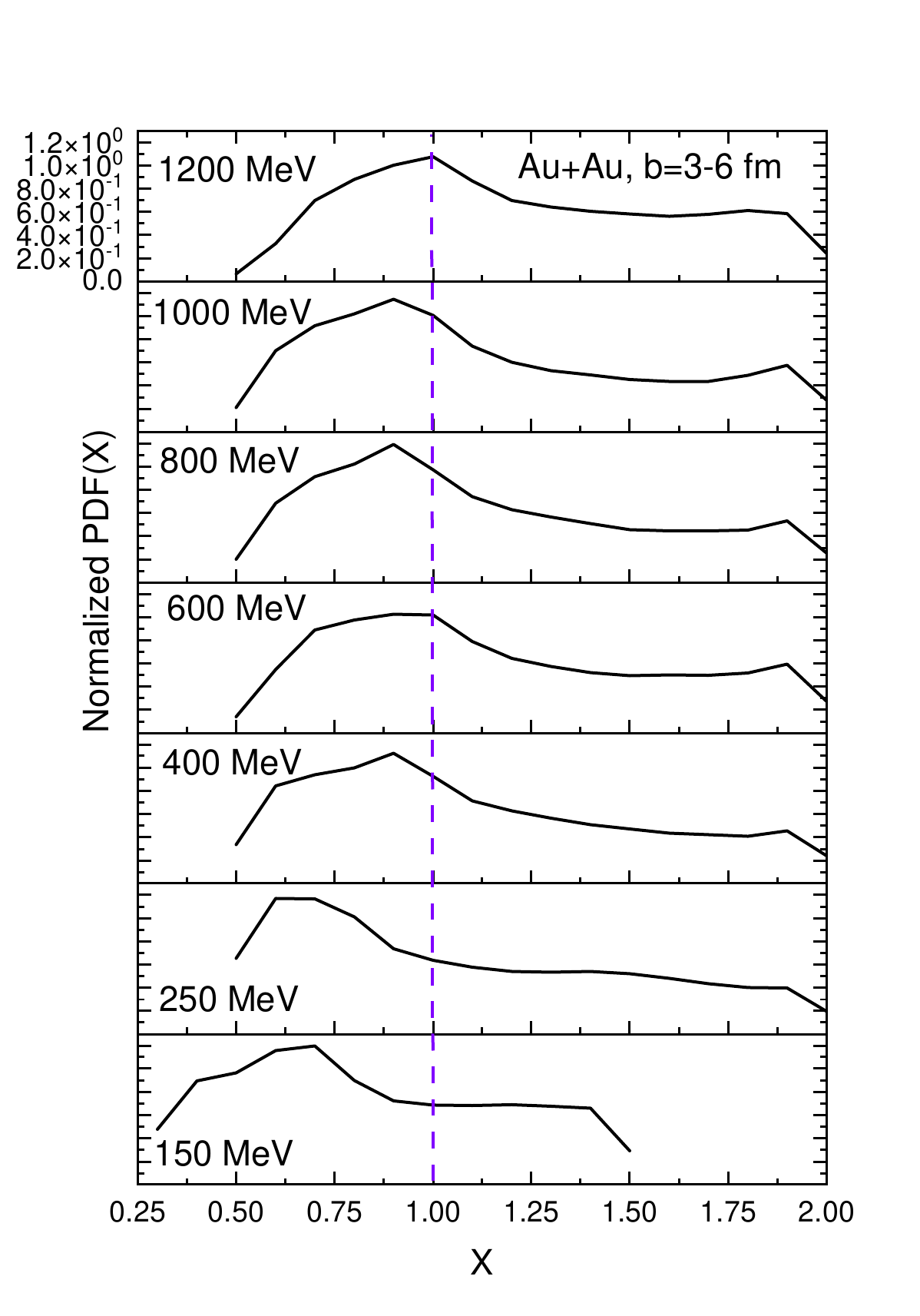}
}
 \resizebox{0.49\textwidth}{!}{
\includegraphics[width=1\textwidth]{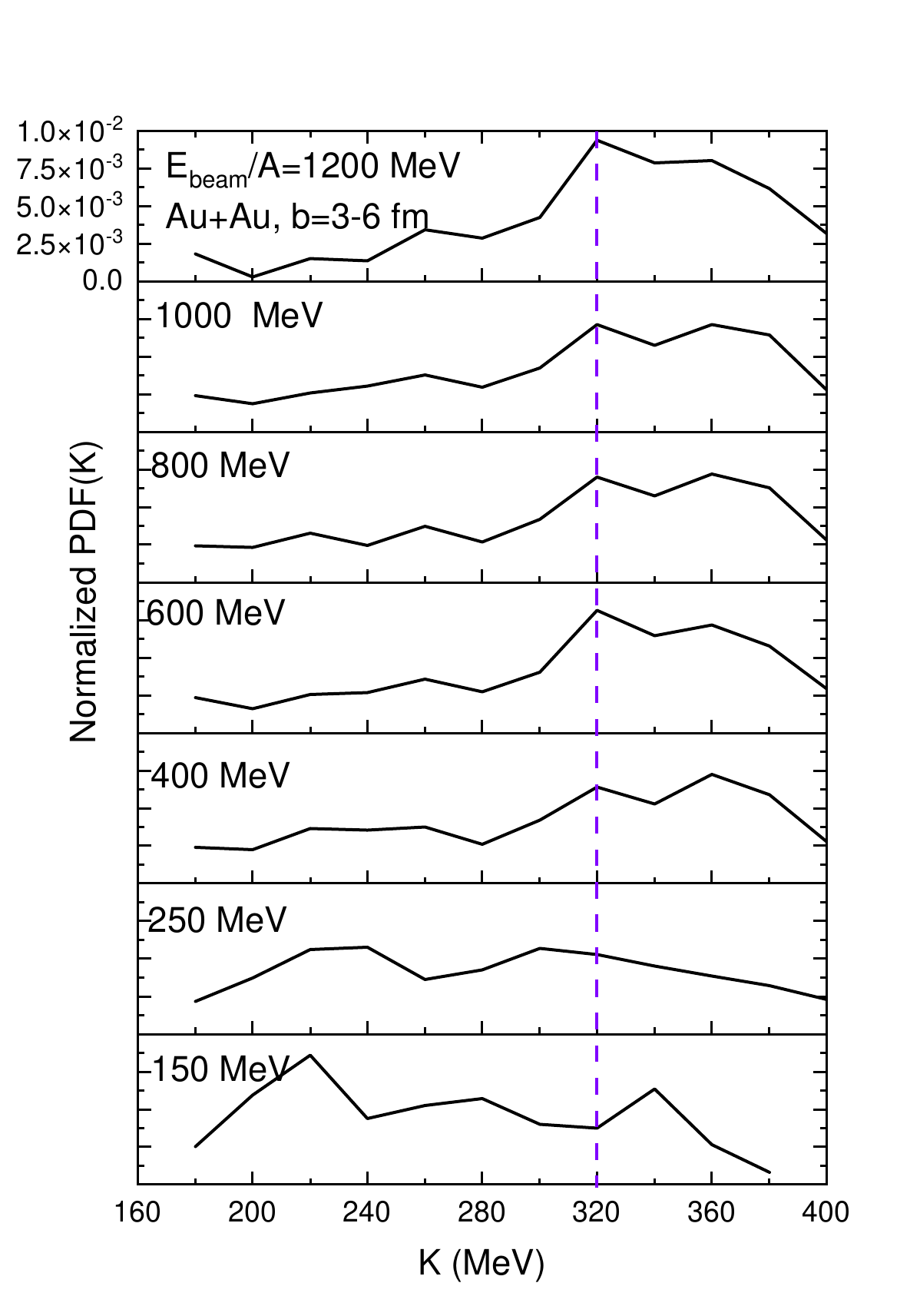}
}
\caption{Beam energy dependence of the posterior PDFs of $X$ (left) and $K$ (right) inferred from the FOPI proton flow data. The vertical dashed lines at $X=1$ and $K=320$ MeV are references. The PDFs are all normalized to 1. For clarity, tick labels are only shown for the top panel.} \label{XK-E}
\end{figure*}
\subsection{Typical posterior PDFs of $X$ and $K$ and their correlations}
As representatives, shown in the upper panels of Figs. \ref{ProfileE150}, \ref{ProfileE250}, and \ref{ProfileE1000} are the $X-K$ correlations (heat maps, also referred as 2D PDFs) and their profiles (on the upper and right sides) along the constant $X$ or $K$ lines indicated from our Bayesian analyses of the FOPI data at a beam energy of 150, 250 and 1000 MeV/nucleon, respectively. The corresponding posterior 1D PDFs (solid) of $X$ and $K$ together with their prior PDFs (dashed rectangles) are shown in the lower panels of each figure. We present the normalized 1D posterior PDFs to make meaningful comparisons with their uniform prior PDFs. The following observations are worth emphasizing:
\begin{enumerate}
    \item With respect to the uniform prior PDF of $X$, there are strong indications that the FOPI data favor reduced in-medium BBSCSs especially at both 150 MeV/nucleon and 250 MeV/nucleon. In particular, the profile of $X$ of the 2D PDFs along the blue lines with constant $K$ values where the PDF(K) has a local or global maximum (most probable value), or more generally and consistently the 1D PDF(X), peaks around 0.7 at both $E_{beam}/A$=150 and 250 MeV, indicating most likely reduced in-medium BBSCSs are necessary to reproduce the data. This peak moves to about $X=0.9$ when the beam energy is increased to 1000 MeV. We caution that because of the long tails in the region $1.0\leq X\leq 2.0$, the mean value of $X$ may still be larger than 1.0 especially at high beam energies. 
    \item The profile of $K$ in the narrow bands bounded by the red vertical lines around the most probable values of $X$ in the heat map, or more generally and consistently the PDF(K), often has multiple local peaks, especially at 150 and 250 Mev/nucleon beam energies. It indicates that the flow data used do not tightly constrain the stiffness of dense matter formed in these reactions. Namely, there are multiple combinations of $X-K$ that can equally well describe the flow data used. For example, at 150 MeV/nucleon beam energy, the PDF(K) has 3 local peaks at $K\approx$ 220 MeV (the most probable one), 280 MeV and 340 MeV, respectively. No wonder it has been so hard to infer the incompressibility from heavy-ion reactions. This finding is consistent with the indications of the original IBUU results (their projections onto the $v_{11}-K$ and $v_{20}-K$ planes spread out in broad ranges of $K$) shown in Figs. \ref{train1}  and \ref{train2}.
  \item Comparing the posterior PDFs of $X$ and $K$ with their respective prior PDFs, it is seen that the $X$ is better constrained although its PDF has long tails with $X\geq 1.0$. This is consistent with the relatively narrow bands of green dots when the $v_{11}$ and $v_{20}$ are projected to the $v_{11}-X$ and $v_{20}-X$ planes as shown in Figs. \ref{train1} and \ref{train2}.

  \item As the beam energy increases from 150 to 1000 MeV/nucleon, the major peak of PDF(K) changes from the most probable value around $K\approx 220$ MeV to $K\approx 320$ MeV, indicating that the overall stiffness of hot and dense matter formed in these reactions has evolved from being relatively soft to stiff. Interestingly, $K\approx 220$ is consistent with the world average of nuclear incompressibility $K_0\approx 220\sim 240$ MeV from studying giant resonances \cite{Blaizot,Garg18,Roca,LiXie21}. As we discussed earlier, it is only meaningful to compare the stiffness parameter $K$ from relatively low-energy heavy-ion reactions with the incompressibility $K_0$ from studying giant resonances. 
  As shown in Fig. \ref{J0Z0K} and discussed in detail earlier, the $K$ parameter characterizes the overall stiffness of the cold nuclear matter in the entire dense region probed by heavy-ion reactions. The observed PDF(K) evolution with beam energy is easily understandable. 
  
  \item Interestingly, at $E_{beam}/A=$250 MeV, the PDF(K) has two major peaks around $K\approx 220$ MeV and $K\approx 320$ MeV having approximately equal heights as if the hot and dense matter formed in this reaction is in a transitional state between a soft and a stiff matter. Interestingly, as shown in Fig. \ref{v12}, the experimental data for $v_{20}$ is quickly decreasing while the $v_{11}$ is raising at this beam energy. It indicates that the driving force for flow is changing very rapidly. It would thus be interesting to investigate in more detail in the future using more observables and possibly with more beam energies in smaller steps around 250 MeV/nucleon. 
\end{enumerate}

\subsection{Evolutions of posterior PDFs of $X$ and $K$ and their correlations with beam energy}

\begin{figure}[thb]
\vspace{-1cm}
\centering
\hspace{-1.cm}
 \resizebox{0.53\textwidth}{!}{
\includegraphics[width=1.\textwidth]{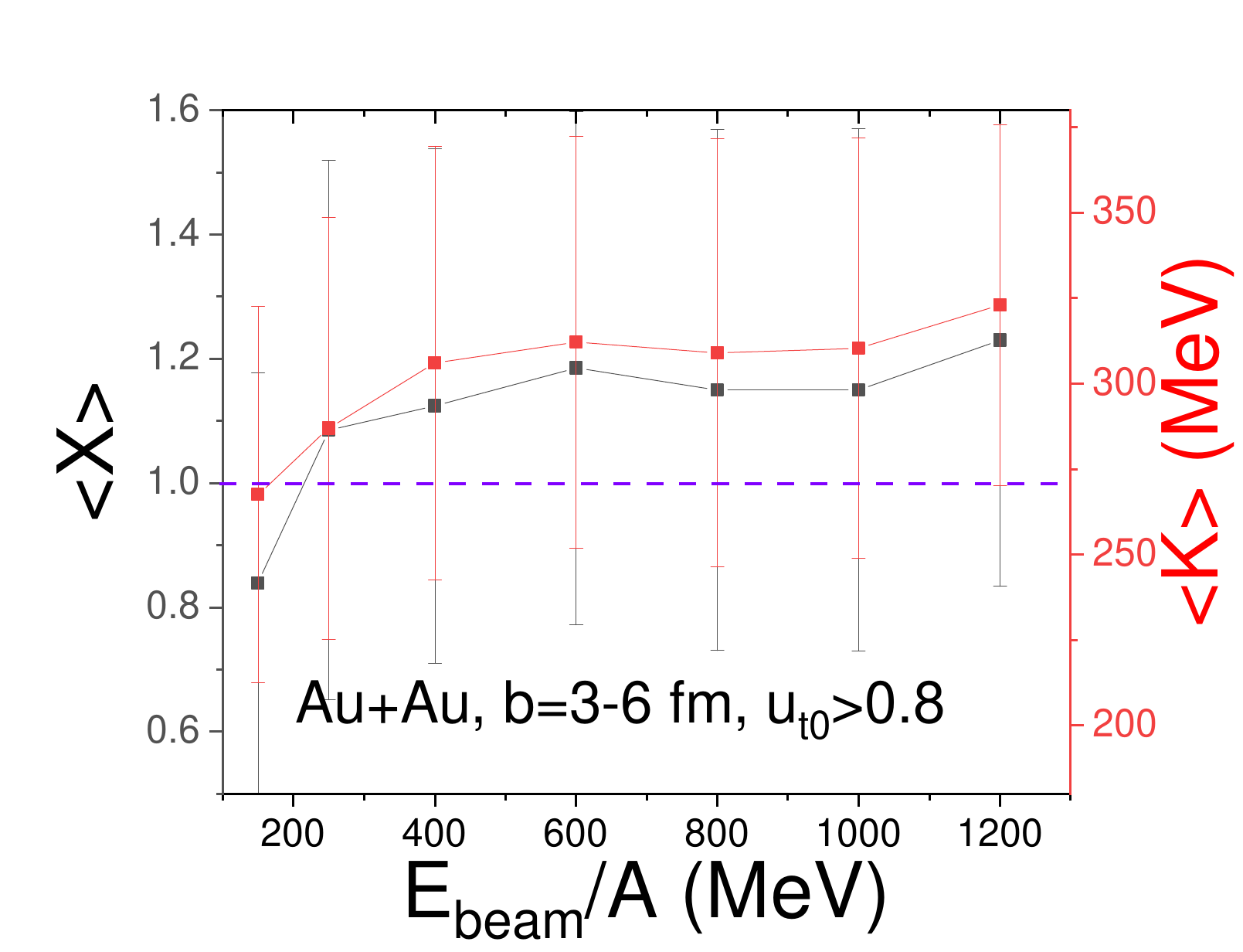}
}
\caption{Mean values of $X$ (left, black) and $K$ (right, red) as functions of beam energy inferred from the FOPI proton flow data.}\label{XK-mean}
\end{figure}

Shown in Fig. \ref{XK-E} are 
the beam energy dependences of the posterior PDFs of X (left) and K (right), respectively. It is seen that the major peak of the posterior PDF of $X$ evolves from $X_{max}\leq 1$ to $X_{max}\approx 1$ with an extended tail at $X\geq 1$ as the beam energy increases from 150 to 1200 MeV/nucleon. On the other hand, the major peak of posterior PDF($K$) evolves from being soft with $K\approx 220$ MeV to stiff with $K\approx 320$ MeV with an extended shoulder in $K\approx 320\sim 370$ MeV. Moreover, one should not expect a unique posterior PDF($K$) peaked at a single value of $K$ in all Au+Au reactions at different beam energies.
Since an increased $X$ generally leads to a higher temperature that also makes matter more stiff, altogether, the FOPI proton flow excitation function data indicate a gradual hardening of hot and dense nuclear matter formed as its density and temperature both increase with higher beam energies. 

It is worth emphasizing here that the increases in $K$ (again, $K$ is not just the incompressibility at $\rho_0$ but also the stiffness of EOS at high densities as illustrated in Fig. \ref{J0Z0K}.) with increasing beam energy simply reflects the hardening of the EOS as the density increases. On the other hand, as we mentioned in Section II B, the community has multiple predictions on why/how the in-medium BBSCS may change with density and temperature. Consistent with findings of earlier work, the most probable value of $X$ generally increases as the beam energy increases, albeit remains less than 1 in the beam energy range studied. It is known that the free-space BBSCS itself decreases and the Pauli blocking becomes less important with increasing beam energy. Our results on the evolution of $X$ indicate that the flow data require increasingly more particle-particle collisions to stop the colliding nuclei and generate enough transverse momenta as the beam energy increases. Thus, the effective BBSCSs (controlled by the $X$ parameter) generally increase with the increasing beam energy.

\begin{table}
\centering
\caption{Beam energy dependence of the mean values of $X$ and $K$ as well as their standard deviations, the MaP of X and its 68\% confidence boundaries inferred from the Bayesian analyses of FOPI's excitation functions of proton directed and elliptic flow data shown in Fig. \ref{v12}.}
\begin{tabular}{cccc}\label{mean}
\\\hline
$E_{\rm{beam}}$ (MeV) &$<X>$ & $<K>$ (MeV) & X (MaP) \\\hline
150 &$0.84\pm 0.34$ &$267.6\pm 55.0$ & $0.7^{+1.1}_{-0.4}$\\
250 &$1.09\pm 0.43$ &$286.9\pm 61.8$& $0.6^{+1.3}_{-0.5}$\\\
400 &$1.12\pm 0.41$ &$306.1\pm 63.4$& $0.9^{+1.3}_{-0.8}$ \\
600 & $1.19\pm 0.41$ &$312.1\pm 60.2$& $0.9^{+1.4}_{-0.8}$\\
800 &$1.15\pm 0.41$ &$309.0\pm 62.7$& $0.9^{+1.3}_{-0.7}$ \\
1000 &$1.15\pm 0.42$ &$310.4\pm 61.5$ & $0.9^{+1.3}_{-0.7}$\\
1200 &$1.23\pm 0.40$ &$323.0\pm 52.8$ & $1.0^{+1.4}_{-0.9}$\\
\hline
\end{tabular}
\vspace{-0.2cm}
\end{table}
To be more quantitative, shown in Fig. \ref{XK-mean} (the corresponding data are given in Table \ref{mean}) are the mean values of $X$ (left, black) and $K$ (right, red) as functions of beam energy. Obviously, there is a transition around $E_{beam}/A$=250 MeV. The stiffness $K$ changes from being soft to stiff,
and the mean value of in-medium BBSCS modification factor $X$ changes flow less to larger than 1. Ironically, the mean value of $K\approx 320$ MeV extracted from Au+Au reactions at beam energies above about 400 MeV/nucleon is consistent with those available in the literature from analyzing heavy-ion reactions using momentum-independent interactions \cite{Pawel02,Fuchs06}, while $K\approx 220$ MeV from analyzing the reaction at 150 MeV/nucleon is consistent with that from studying giant resonances as we mentioned earlier. 

\begin{figure}[thb]
\vspace{-1cm}
\centering
\hspace{-1.cm}
 \resizebox{0.53\textwidth}{!}{
\includegraphics[width=1.\textwidth]{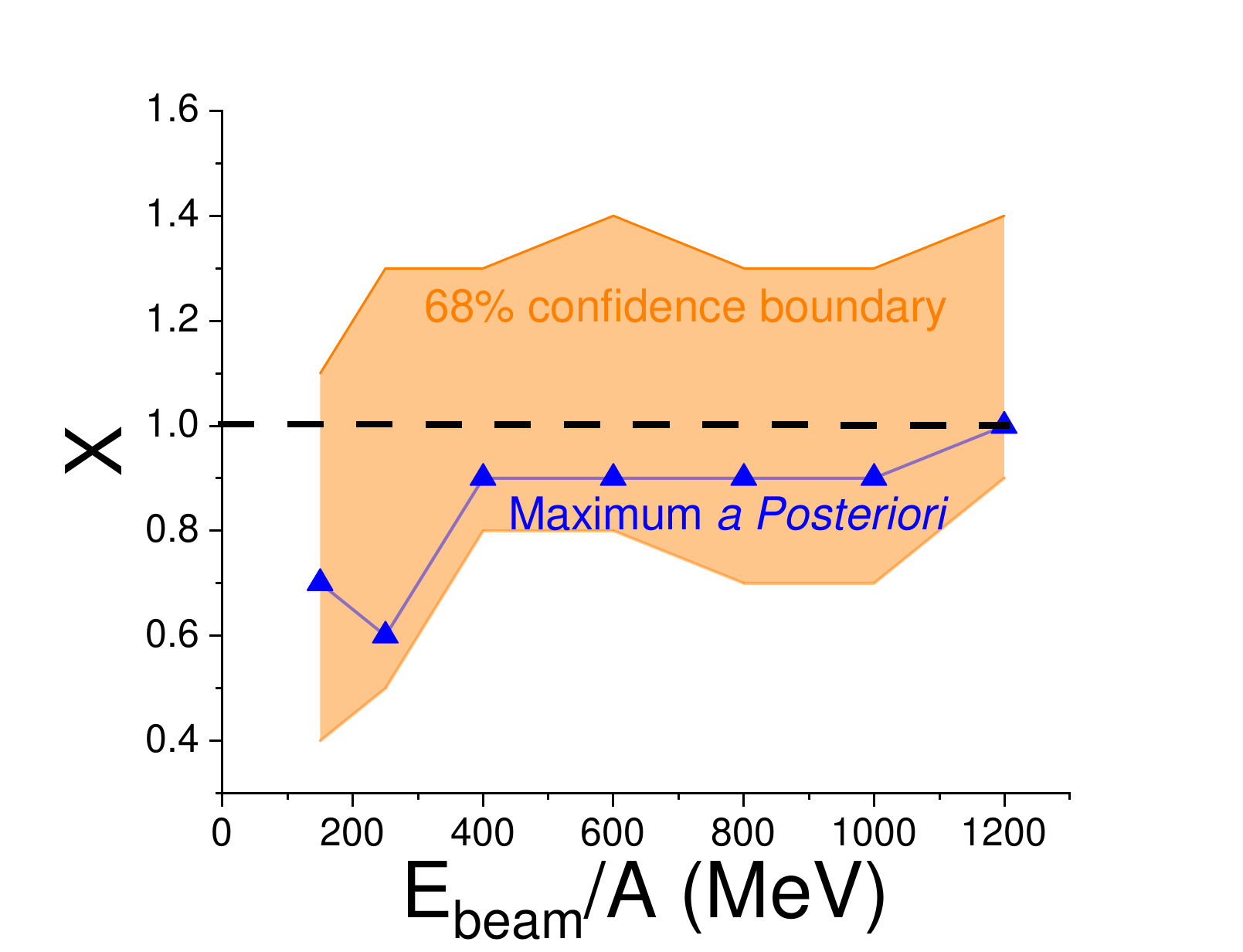}
}
\caption{The MaP value of $X$ as a function of beam energy inferred from the FOPI proton flow data.}\label{X-map}
\end{figure}

Compared to the detailed information embedded in the posterior PDFs of $X$ and $K$, their mean values only provide some rough information regarding their general trends. In particular, the general trend of $<X>$ is qualitatively consistent with existing knowledge in the literature. However, quantitatively, the mean values of $X$ we obtained are more close to the upper end of its current uncertainty range. We emphasize that the mean $X$ is strongly influenced by its prior upper limit. As mentioned earlier, the flow data used do not constrain strongly enough the $X$ to make its posterior PDF to vanish at its prior upper boundary (although it is strongly reduced compared to its uniform prior PDF as discussed above). If one sets the prior upper limit of $X$ to be less than $2$ that we used here, the resulting posterior PDF of $X$ would then have a shorter high-X tail that will make the mean $<X>$ smaller correspondingly. Nevertheless, considering inclusively the diversity of existing predictions on X in the literature as discussed in Section \ref{Pri}, it is fair and useful to set the prior upper limit of $X$ at 2 for the stated purposes of this work. However, we caution the reader that our results for $<X>$ might suffer from a systematic error favoring higher $<X>$ depending on how one sets the prior range of $X$. As we mentioned earlier, including more observables may help remove or reduce this uncertainty.  

We point out that the above results are consistent with our previous finding of $<X>=1.32^{+0.28}_{-0.40}$ and $<K>=346^{+29}_{-31}$ from analyzing the proton flow data taken by the HADES Collaboration for the mid-central Au+Au reaction at $E_{beam}/A$=1.23 GeV. It is also interesting to note that there are flow data at beam energies below 150 MeV/nucleon from other experiments using different detectors, we plan to extend our Bayesian analyses to those data in the near future. 

For a comparison and to be more accurate in interpreting our results, shown in the last column of Table \ref{mean} and 
Fig. \ \ref{X-map} are the Maximum {\it a Posteriori} (MaP) and 68\% confidence boundaries of the posterior PDF($X$). The MaP itself is expected to be less influenced than its confidence boundaries by the uncertainties in setting the prior upper limit of $X$. Consequently, the MaP of $X$ is less affected than its mean $<X>$ by the possible biases in setting its prior range. It is worth noting that because of the dual or more peaks with compatible heights in the posterior PDF of $K$, it is not practically very useful to discuss the MaP of $K$ in this work. Following the procedure and techniques given in Ref. \cite{Turkkan14}, the lower $X_{\mathrm{L}}$ and upper $X_{\mathrm{U}}$ 68\% confidence boundaries of the highest posterior density (HPD) 
for $X$ are found from examining the integral
\begin{equation}\label{HPD}
  \int_{X_{\mathrm{L}}}^{X_{\mathrm{U}}}\mathrm{PDF}(X)dX=0.68.
\end{equation}
Reflecting the high-X tail in the very asymmetric posterior PDF(X), the $X_{\mathrm{L}}$ and $X_{\mathrm{U}}$ are highly asymmetric with respect to the MaP of $X$. Interestingly, the latter increases from about 0.6 at $E_{\rm beam}/A$ around 200 MeV/A to about 0.9 from $E_{\rm beam}/A$ between 400 and 1000 MeV/A, before reaching about 1 at 1200 MeV/A. This result is more consistent quantitatively with the existing knowledge on $X$ from forward-modelings of collective flow in the literature compared to what we learned above from examining the beam energy dependence of $<X>$.

It is also worth emphasizing that the interpretation of results from Bayesian analyses has to be done with respect to the prior ranges and PDFs used. For the present analyses of the FOPI data, the resulting improvement of our knowledge regarding the $X$ and $K$ parameters by analyzing the FOPI data is impressive with respect to their broad and uniform prior PDFs. However, it is clearly seen that the posterior PDFs of $X$ and $K$ are not zero at the upper and lower boundaries of their prior PDFs. Of course, technically one can remove this by conducting analyses in even larger prior ranges that is computationally costly. Physically, however, we believe it indicates limitations of the selected data we used in constraining their posterior PDFs. Consequently, the standard deviations of $<X>$ ($<K>$) as well as the tails (shoulders) in the posterior PDFs of $X$ ($K$) are still large. In our opinion, these are mostly due to our decision of using only the $v_{11}$ and $v_{20}$ data in mid-central Au+Au reactions at each beam energy for the stated purposes of this work. In fact, both FOPI and HADES Collaboration have published data of many differential (e.g., rapidity and transverse momentum dependence) observables and their dependences on centrality using different triggers (e.g., cuts on transverse momentum). Incorporating these differential observables and their impact parameter dependences in more comprehensive Bayesian analyses are expected to provide more stringent constrains on the posterior PDFs of model parameters, albeit requiring much more computing time and human efforts to train their GP emulators that we can not afford at this time. Nevertheless, the general conclusions of our present work are expected to stay. Testing these expectations is on our future working plan.  

\section{Summary, caveats and outlook}\label{sum}
In summary, within the Bayesian statistical framework using the GP emulator for the IBUU transport model simulators 
with momentum-independent Skyrme interactions, we inferred from the proton directed and elliptical flow data in mid-central Au+Au reactions at beam energies from 150 to 1200 MeV/nucleon the evolutions of in-medium BBSCS modification factor $X$ and the stiffness parameter $K$ of dense nuclear matter. We found that the most probable value of $X$ evolves from around 0.7 to 1.0 as the beam energy $E_{beam}/A$ increases. On the other hand, the posterior PDF($K$) may have dual peaks having roughly the same height or extended shoulders at high $K$ values. More quantitatively, the posterior PDF($K$) changes from having a major peak around 220 MeV characterizing a soft EOS in the reaction at $E_{beam}/A$=150 MeV to one that peaks around 320 MeV indicating a stiff EOS in the reactions at $E_{beam}/A$ higher than about 600 MeV. The transition from soft to stiff happens in mid-central Au+Au reactions at beam energies around 250 MeV/nucleon in which $K=220$ MeV and $K=320$ MeV are approximately equally probable. Consequently, the 
means of $X$ and $K$ change from $<X>=0.84\pm 0.34$ \& $<K>=267.6\pm 55.0$ MeV at $E_{beam}/A$=150 MeV to $<X>=1.23\pm 0.40$ \& $<K>=323.0\pm 52.8$ MeV at $E_{beam}/A$=1200 MeV. We conclude that the FOPI proton flow excitation function data indicate a gradual hardening of the EOS of hot and dense nuclear matter as its density and temperature increase in the reactions with higher beam energies.

Certainly, our work presented here has limitations and caveats. Chiefly among them, we notice the following
\begin{enumerate}
    \item We do not and can not claim our results are model independent. The authors of this work and the IBUU code used are participants of the Transport Model Evaluation Project (TMEP). As discussed extensively in the recent TMEP review \cite{Herman} and references therein, besides different inputs and modelings of many microscopic processes during heavy-ion collisions, there are also technical differences in realizing them in various transport models. Consequently, some observables predicted by different transport models still suffer from certain model dependence. The differences among transport model predictions are up to about 15\% for most nucleonic observables but larger for mesonic ones. For example, the proton flow observables and the inferred posterior PDFs of the two model parameters ($X$ and $K$) certainly depend on the efficiency of the Pauli-blocking implemented in the IBUU during the whole evolution dynamic process. Thus, our conclusions should all be understood within the context of using the IBUU transport code. As always, the readers should be aware of possible model dependence in interpreting results obtained from essentially all transport modes and some other types of models for heavy-ion reactions. Moreover, to our best knowledge, presently there is no practical way to evaluate convincingly and reliably systematic errors in transport model predictions.
    \item Using the minimum net $\sigma_{\mathrm{obs},j}$ in the likelihood function is the best-case scenario. While it allowed us to evaluate the maximum constraining power of the FOPI flow data on the PDFs of (X and K) model parameters, we certainly have underestimated their posterior uncertainties. Future work should explore ways to first realistically evaluate and then incorporate properly both statistical and systematical errors of simulations and emulations. However, since the posterior uncertainties of both $X$ and $K$ are already rather big (although much better than their prior PDFs) by using only the minimum value of $\sigma_{\mathrm{obs},j}$, just incorporating more sources of errors in the likelihood function is unlikely to help improve anything unless more experimental data of relevant observables (identifying such observables sensitive to the specified model parameters has been a challenging task for the heavy-ion reaction community for several decades) are used in more comprehensive Bayesian analyses.
\end{enumerate}

Looking forward, what aspects of our approach should be improved first such that some of the caveats mentioned above may be removed quickly? Immediately coming to our mind is to couple the IBUU code with a reasonable coalescence model for identifying light clusters. This will enable us to use the vast amount of data taken for cluster observables in many experiments. Applying the simplest coalescence model at the final state of heavy-ion collisions requires two additional model parameters. While Bayesian analyses using simulators with multiple model parameters and emulators for multiple observables are costly, the potentially strong scientific rewards may be appealing for significantly more investments in them. Hopefully, our work presented here serves as a useful starting point for more efforts in this direction.

\section*{Acknowledgement}
We thank Xavier Grundler for helpful discussions. 
BALI was supported in part by the U.S. Department of Energy, Office of Science,
under Award No. DE-SC0013702, the CUSTIPEN (China-
U.S. Theory Institute for Physics with Exotic Nuclei) under
US Department of Energy Grant No. DE-SC0009971. WJX is supported in part by the Open Project of Guangxi Key Laboratory of Nuclear Physics and Nuclear Technology, No. NLK2023-03 and the Central Government Guidance Funds for Local Scientific and Technological Development, China (No. Guike ZY22096024).



\end{document}